\documentclass[aps,showpacs,amsmath,amssymb,twocolumn,pra,superscriptaddress,notitlepage]{revtex4-2}

\usepackage{qcircuit}
\usepackage{amsmath,bm}
\usepackage[dvips]{graphicx}
\usepackage{amsmath,amssymb,amsthm,mathrsfs,amsfonts,dsfont}
\usepackage{subfigure, epsfig}
\usepackage{braket}
\usepackage{bm}
\usepackage{enumerate}
\usepackage{color}
\usepackage{graphicx}
\usepackage{algorithm}
\usepackage{algorithmic}
\usepackage{braket}
\usepackage{comment}
\usepackage{appendix}
\usepackage{here}
\usepackage{tabularx}
\usepackage{physics}
\usepackage{mathtools}

\newcommand{\red}[1]{\textcolor{black}{#1}}

\begin{document}


\title{Unitary-transformed projective squeezing: applications for circuit-knitting and state-preparation of non-Gaussian states }


\author{Keitaro Anai}
\affiliation{Department of Applied Physics, School of Engineering, The University of Tokyo,
7-3-1 Hongo, Bunkyo-ku, Tokyo 113-8656, Japan}
\affiliation{NTT Computer and Data Science Laboratories, NTT Corporation, Musashino, Tokyo 180-8585, Japan}

\author{Yasunari Suzuki}
\email{yasunari.suzuki@ntt.com}
\affiliation{NTT Computer and Data Science Laboratories, NTT Corporation, Musashino, Tokyo 180-8585, Japan} 

\author{Yuuki Tokunaga}
\affiliation{NTT Computer and Data Science Laboratories, NTT Corporation, Musashino, Tokyo 180-8585, Japan} 

\author{Yuichiro Matsuzaki}
\affiliation{Department of Electrical, Electronic, and Communication Engineering, Faculty of Science and Engineering, Chuo University, 1-13-27 Kasuga, Bunkyo-ku, Tokyo 112-8551, Japan} 

\author{Shuntaro Takeda}
\affiliation{Department of Applied Physics, School of Engineering, The University of Tokyo,
7-3-1 Hongo, Bunkyo-ku, Tokyo 113-8656, Japan}

\author{Suguru Endo}
\email{suguru.endou@ntt.com}
\affiliation{NTT Computer and Data Science Laboratories, NTT Corporation, Musashino, Tokyo 180-8585, Japan}


\begin{abstract}
Continuous-variable (CV) quantum computing is a promising candidate for quantum computation because it can, even with one mode, utilize infinite-dimensional Hilbert spaces and can efficiently handle continuous values. Although photonic platforms have been considered as a leading platform for CV computation, hybrid systems that use both qubits and bosonic modes, e.g., superconducting hardware, have shown significant advances because they can prepare non-Gaussian states by utilizing the nonlinear interaction between the qubits and the bosonic modes. However, the size of hybrid hardware is currently restricted. Moreover, the fidelity of the non-Gaussian state is also restricted. This work extends the projective squeezing method to establish a formalism for projecting quantum states onto the states that are unitary-transformed from the squeezed vacuum at the expense of the sampling cost. Based on this formalism, we propose methods for simulating larger quantum devices and projecting states onto the cubic phase state, a typical non-Gaussian state, with a higher squeezing level and higher nonlinearity. To make implementation practical, we can, by leveraging the interactions in hybrid systems of qubits and bosonic modes, apply the smeared projector by using either the linear-combination-of-unitaries or virtual quantum error detection algorithms. We numerically verify the performance of our methods and show that projection can suppress the effect of photon-loss errors.

\end{abstract}

\maketitle
\section{Introduction}
Due to the recent significant advances in hardware and algorithms, quantum computing is now the subject of unprecedented anticipation for the realization of practical quantum computing in diverse fields~\cite{cerezo2021variational,montanaro2016quantum}, e.g., quantum machine learning~\cite{mitarai2018quantum,biamonte2017quantum}, and quantum simulations for chemistry and condensed matter~\cite{georgescu2014quantum,bauer2020quantum}, etc. While the qubit-based quantum computing paradigm is receiving the most attention, continuous-variable (CV) quantum computation offers a potentially hardware-efficient route to the realization of quantum computers by using the infinitely large Hilbert space of bosonic systems~\cite{adesso2014continuous,cai2021bosonic,joshi2021quantum,ths:CV-QAOA,ths:CV-QML}. With regard to realizing universal CV quantum computing, it has been shown that an arbitrary unitary operators yielded by a polynomial number of bosonic operators can be constructed with Gaussian operations and one type of non-Gaussian operation, e.g., a cubic phase gate (CPG)~\cite{lloyd1999quantum}.

Although photonic systems were initially considered to be leading candidates for CV computing due to their scalability~\cite{ths:Rev_Takeda}, the absence of sufficient non-Gaussianity renders universal quantum computing challenging. However, hybrid architectures consisting of two-level systems and bosonic modes have been developed to realize various non-Gaussian gate operations~\cite{cai2021bosonic,joshi2021quantum, sivak2023real,eickbusch2022fast,de2022error,kudra2022robust}. \textcolor{black}{Nevertheless, the scale of the experiments with this type of hardware is currently limited because of leveraging matter systems such as superconducting circuits and ions, with the number of control elements increasing as the system size system expands.} In addition, while the cubic phase state (CPS) has recently been demonstrated in hybrid systems ~\cite{kudra2022robust,eriksson2024universal}, the achievable fidelity remains around $92\%$ ~\cite{eriksson2024universal}.



\begin{figure}
    \centering
    \includegraphics[width=1.\linewidth]{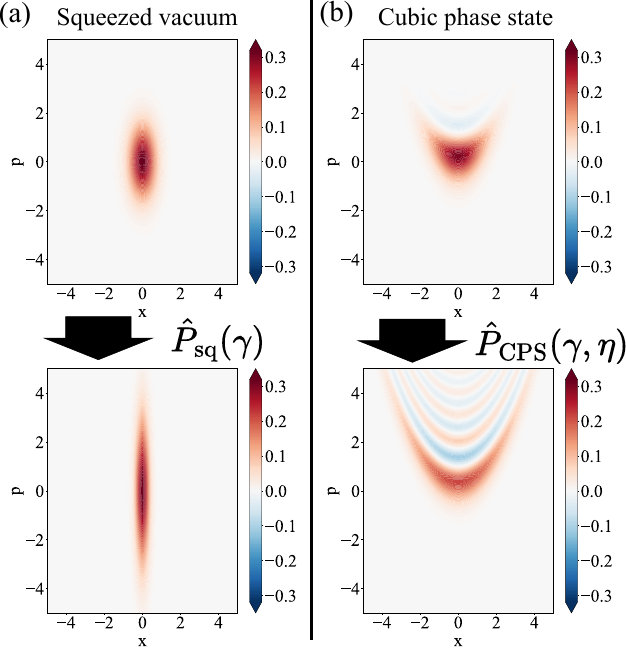}
    \caption{Wigner functions of single-mode states before and after (unitary-transformed) projective squeezing. $\hat{P}_\mathrm{sq}(\gamma)$ and $\hat{P}_\mathrm{CPS}(\gamma,\eta)$ are smeared projectors onto the state with a higher squeezing level of the squeezed vacuum and CPS, respectively. Parameter $\gamma$ determines the increase in squeezing level as introduced in the main text. Here, we consider a 3-dB squeezed vacuum and the CPS developed from a 3-dB squeezed vacuum as initial states and choose parameter $\gamma$ of the smeared projector to increase the squeezing level by 3 dB. (a) Wigner functions of the squeezed vacuum before and after projective squeezing. After projective squeezing, the width of the squeezed vacuum decreases; the squeezing level of the squeezed vacuum rises. (b) Wigner functions of the CPS before and after projective squeezing. After projective squeezing, the stripes of the Wigner function sharpen.}
    \label{fig:Wigner}
\end{figure}

In the present work, by extending the recently introduced \textit{projective squeezing} method~\cite{endo2024projective}, we propose a protocol for CV quantum computing that can project the state onto the subspace of interest to compensate resource paucity, i.e., entanglement for large-scale CV computation and non-Gaussianity for universality, at the cost of higher sampling overhead and controlled operations on ancilla qubits. The \textit{projective squeezing } method allows the squeezing level of a squeezed vacuum state to be increased by applying the smeared projector with a linear combination of displacement operators with Gaussian weight~\cite{endo2024projective}. This is possible because the displacement operators $\{\hat{D}(\alpha) \}_\alpha ~ (\alpha \in \mathbb{C})$ towards the anti-squeezing axis constitute the stabilizers of an infinitely squeezed vacuum state. Accordingly, we propose \textit{unitary-transformed projective squeezing} by utilizing the fact that useful resource states can be connected to single-mode squeezed vacuum states or the tensor product of squeezed vacuum states through unitary operator $\hat{U}$. In this case, the stabilizers of these resource states can be written as $\hat{V}(\alpha)=\hat{U} \hat{D}(\alpha) \hat{U}^\dag $. Hence, we can construct the smeared projector for the target states with $\hat{V}(\alpha)$. Figure~\ref{fig:Wigner} illustrates projective squeezing and unitary-transformed projective squeezing. We show that our method can be performed either with the linear-combination-of-unitaries (LCU) algorithm~\cite{ths:LCU_1,ths:LCU_2}, which physically post-selects the quantum state transformed by the linear combination of unitary operations, or virtual quantum error detection (VQED)~\cite{ths:VQED}, which allows the expectation values corresponding to the projected quantum states to be computed. Note that we can perform the LCU algorithm for Gaussian coefficients with the iteration of a controlled operation using a single ancilla qubit~\cite{endo2024projective}. Because the VQED method also necessitates controlled operations, our framework requires controlled-$\hat{V}(\alpha)$ operation for projection.

As practical applications of our method, we show that we can project states onto two types of quantum states: entangled states and non-Gaussian states. First, we show that the stabilizers for the CV Einstein-Podolsky-Rosen (EPR) and cluster states, the typical CV entangled states, are tensor products of local displacement operators. We reveal that the local VQED implementation of the distillation of cluster state with local controlled-displacement operations from separable ancilla qubits allows simulation of larger size quantum computations, i.e., so-called circuit knitting~\cite{ths:knitting_1,ths:knitting_2,ths:knitting_3,piveteau2023circuit,harada2023doubly}. This approach ehances the scalability of CV noisy intermediate-scale quantum (NISQ) hardware~\cite{hillmann2020universal,eriksson2024universal}. Second, we can also project the states onto CPS, one of the typical CV non-Gaussian states, by revealing that the stabilizers are the product of squeezing, displacement, and phase-shift operators, all of which are Gaussian operations. This leads to improvements in the non-Gaussianity for universal computing in CV quantum computing. We note that several existing theoretical schemes for non-Gaussian states often rely on higher-order Hamiltonians, such as cubic or quartic terms. In contrast, our method requires only Gaussian operations (generated by up to second-order Hamiltonians) and controlled phase-shift gates, which have been demonstrated experimentally with relatively low overhead. By conducting numerical simulations and verifying the performance of our protocol, we also show that projection onto CPS and the cluster state can mitigate the effect of photon-loss errors.

Our framework can be implemented with experimental platforms with flexible controlled operations. In particular, superconducting hardware offers significant varieties of controlled operations due to the flexible design of the superconducting circuit and the strong interaction between physical systems~\cite{cai2021bosonic,joshi2021quantum}. The controlled displacement operations required for circuit knitting have been experimentally demonstrated as well as being used in the stabilization of GKP qubits~\cite{eickbusch2022fast,sivak2023real}. The controlled squeezing operations required for projection onto CPS were also proposed for superconducting hardware in Refs.~\cite{ayyash2024driven,del2024controlled}.





The rest of this paper is organized as follows. In Sec \ref{sec:ProjectiveSqueezing}, we review the recently proposed projective squeezing method, and focus on increasing the squeezing level of the squeezed vacuum state. In Sec.~\ref{sec:Unipro}, we discuss the unitary-transformed projective squeezing that allows for projection to the more general states that can be obtained by transforming the single-mode squeezed vacuum state or the tensor products of the squeezed vacuum states with unitary operators. Then, we introduce practical applications of this framework for circuit knitting and projection onto CPS. In Sec.~\ref{sec: Physimp}, after we review the LCU and VQED methods, we describe the physical implementation of circuit knitting and the projection onto CPS. Sec.~\ref{sec:Simulations} details numerical simulations of our protocol. We finally conclude our paper with discussions and conclusions.

\label{section: introduction}





\section{Projective squeezing}\label{sec:ProjectiveSqueezing}

Here, we review the projective squeezing method for the squeezed vacuum state~\cite{endo2024projective}. Let us denote the bosonic annihilation and creation operators as $\hat{a}$ and $\hat{a}^\dagger$, respectively; they satisfy the commutation relation $[\hat{a}, \hat{a}^\dagger]=1$. Then, position and momentum quadrature operators are described as $\hat{x}=(\hat{a}+\hat{a}^\dagger)/\sqrt{2}$ and $\hat{p}=-i(\hat{a}-\hat{a}^\dagger)/\sqrt{2}$. The displacement, squeezing, and phase-shift operators are defined as $\hat{D}(\alpha)=\exp(\alpha\hat{a}^\dagger-\alpha^\ast\hat{a})$, $\hat{S}(r)=\exp\left(r(\hat{a}^2-(\hat{a}^\dagger)^2)/2\right)$, and $\hat{R}(\phi)=\exp(i\phi\hat{a}^\dagger\hat{a})$, respectively, where $\alpha\in\mathbb{C}$ represents the amount of displacement in the phase space, $r\in\mathbb{R}$ represents the squeezing level, and $\phi\in\mathbb{R}$ is the amount of phase shift. Then, the $x$-squeezed vacuum state can be written as $\ket{\mathrm{sq}_r}=\hat{S}(r)\ket{0}$. At the limit of $r\rightarrow\infty$, the $x$-squeezed vacuum state reaches state $\ket{x=0}$, which is the eigenvector of position quadrature operator $\hat{x}$ with the zero eigenvalue.\\

Following the symmetry expansion technique, which constructs the projector to a symmetric state by a linear combination of the stabilizers of the target subspace~\cite{mcclean2020decoding,ths:SymmetryExpansion}, we can increase the squeezing level of the squeezed vacuum state. More concretely, to increase the squeezing level of the $x$-squeezed vacuum state, we can use the linear combination of the displacement operator on the $p$-axis as a smeared projector because the displacement operator in $p$-axis $\hat{D}(ip_0/\sqrt{2})\ (p_0\in\mathbb{R})$ is a stabilizer of the $x$-squeezed vacuum state. The smeared projector is defined as
\begin{equation}
\hat{P}_\mathrm{sq}(\gamma)=\int\mathrm{d}p_0\sqrt{\frac{\gamma}{\pi}}\exp(-\gamma p_0^2)\hat{D}\left(i\frac{p_0}{\sqrt{2}}\right),
\label{Eq: smearedpro}
\end{equation}
where $\gamma> 0$ determines the degree of improvement in the squeezing level. We can show that 
\begin{equation}
\begin{aligned}
\hat{P}_\mathrm{sq}(\gamma)&=\mathrm{exp}\bigg[-\frac{\hat{x}^2 }{4\gamma}\bigg] \\
&= \int \mathrm{d}x_0 e^{-\frac{x_0^2}{4 \gamma} } \ket{x=x_0}\bra{x=x_0}, 
\label{Eq: projectorsq}
\end{aligned}
\end{equation}
which indicates that $\hat{P}_\mathrm{sq}(\gamma)$ exactly works as a projector to $\ket{x=0}$ for $\gamma \rightarrow +0$. The smeared projector $\hat{P}_\mathrm{sq}$ transforms the $x$-squeezed vacuum state as
\begin{align}
    \hat{P}_\mathrm{sq}(\gamma)\ket{\mathrm{sq}_r}=e^{-\Delta r/2}\ket{\mathrm{sq}_{r+\Delta r}},\label{eq:VirtualSq_1}\\
    \Delta r=\frac{1}{2} \ln\left(1+\frac{1}{2\gamma e^{2r}}\right).
    \label{eq:VirtualSq_2}
\end{align}
The derivation of Eqs.~\eqref{Eq: projectorsq},~\eqref{eq:VirtualSq_1},~\eqref{eq:VirtualSq_2} is given in Appendix~\ref{sec:ProjSq_SqVac}. From Eqs.~\eqref{eq:VirtualSq_1},~\eqref{eq:VirtualSq_2}, we can increase the squeezing level by $\Delta r$. We see that the projection probability $q_{\Delta r}$, which determines the sampling overhead of this method, can be calculated as
\begin{equation}
q_{\Delta r} = \bra{\mathrm{sq}_r} \hat{P}_\mathrm{sq}(\gamma)^\dag \hat{P}_\mathrm{sq}(\gamma)\ket{\mathrm{sq}_r} = e^{- \Delta r}.
\label{Eq: propro}
\end{equation}
Note that we can also construct the smeared projector that approximately projects the state to $\ket{p=0}$, which is an eigenvector of momentum quadrature operator $\hat{p}$ with the zero eigenvalue, by changing the direction of the displacement towards the $x$-axis. We denote this smeared projector as $\hat{P}_{\mathrm{asq}}(\gamma)$.

We also introduce a nullifier for the squeezed vacuum state because it is useful for evaluating the performance of our protocol. Here, we refer to operator $\hat{\delta}$ as the nullifier of the ideal state $\ket{\psi}$ if equation $\hat{\delta}\ket{\psi}=0$ holds. As the noisy state approaches the ideal state, the variance of the nullifier decreases, a characteristic that can be used for state evaluation. For example, the nullifier of the ideal $x$-squeezed vacuum state with infinite squeezing level is $\hat{x}$ because $\hat{x}\ket{x=0}=0$. For the noisy squeezed vacuum state with finite squeezing level $r$, the variance of the nullifier is given by $\exp(-2r)/2$. The derivation of this relationship is described in Appendix~\ref{sec:Nullifiers}. If we increase the squeezing level in a projective manner by $\Delta r$, the variance of the nullifier turns into $\exp(-2(r+\Delta r))/2$, which approaches zero as $\Delta r$ increases.


\section{Unitary-transformed projective squeezing}
\label{sec:Unipro}
Here, we show that we can project the state onto a unitary-transformed state from an $x$-squeezed state manifold by transforming the smeared projector with a unitary operator. We then discuss its applications: circuit knitting, entanglement enhancement, and preparation for higher-quality CPS. 

\subsection{General formulation}\label{subsec:GeneralFormulation}
Here, we introduce the class of the smeared projector transformed by unitary operator $\hat{U}$ as $\hat{U} \hat{P}_{\rm sq} (\gamma) \hat{U}^\dag$. Since we have
\begin{equation}
\begin{aligned}
&\hat{U} \hat{P}_\mathrm{sq}(\gamma) \hat{U}^\dag= \mathrm{exp}\bigg[-\frac{(\hat{U} \hat{x} \hat{U}^\dag)^2}{4 \gamma} \bigg] \\
&=\int dx_0 e^{\frac{-x_0^2}{4 \gamma }} \hat{U} \ket{x=x_0}\bra{x=x_0} \hat{U}^\dag, 
\label{Eq: general1}
\end{aligned}
\end{equation}
we can see that the unitary-transformed smeared projector exactly acts as projector to state $\hat{U} \ket{x=0}$ for $\gamma \rightarrow +0$. We can also consider the two-mode unitary-transformed smeared projector. For example, 
\begin{equation}
\begin{aligned}
&\hat{U} [\hat{P}_{\rm sq} (\gamma) \otimes \hat{P}_{\rm asq} (\gamma)] \hat{U}^\dag = \mathrm{exp}\bigg[-\frac{\hat{U}(\hat{x}_1^2 +\hat{p}_2^2)\hat{U}^\dag }{4 \gamma} \bigg] 
\label{Eq: twomode}
\end{aligned}
\end{equation}
exactly acts as a projector for state $\hat{U} \ket{x=0}_A \otimes \ket{p=0}_B$ for $\gamma \rightarrow +0$.

To apply the unitary-transformed smeared projector $\hat{U} \hat{P}_\mathrm{sq}(\gamma) \hat{U}^\dag$, we need to expand it with easy-to-implement operators. For example, because $\hat{U}\hat{P}_\mathrm{sq} (\gamma) \hat{U}^\dag$ can be written as

\begin{equation}
\hat{U}\hat{P}_\mathrm{sq} (\gamma) \hat{U}^\dag=\int\mathrm{d}p_0\sqrt{\frac{\gamma}{\pi}}\exp(-\gamma p_0^2) \hat{U} \hat{D}\left(i\frac{p_0}{\sqrt{2}}\right) \hat{U}^\dag,
\label{Eq: unitarysmearedpro}
\end{equation}
we can see that $\hat{U}\hat{P}_\mathrm{sq} (\gamma) \hat{U}^\dag$ is expanded by $\left\{ \hat{U} \hat{D}\left(i\frac{p_0}{\sqrt{2}}\right) \hat{U}^\dag\right\}_{p_0}$. Note that when $\hat{U}$ is a Gaussian operation, $\left\{ \hat{U} \hat{D}\left(i\frac{p_0}{\sqrt{2}}\right) \hat{U}^\dag\right\}_{p_0}$ is also a set of displacement operations.

When the input state is given by $\hat{U} \ket{\mathrm{sq}_{r}}$ with the unitary-transformed smeared projector being applied, we get 
\begin{equation}
\hat{U}\hat{P}_\mathrm{sq} (\gamma) \hat{U}^\dag (\hat{U} \ket{\mathrm{sq}_{r}}) = e^{-\Delta r/2 } \hat{U} \ket{\mathrm{sq}_{r+\Delta r}}.
\end{equation}
Therefore, the projection probability is invariant under unitary transformation,  and the state changes into $\hat{U} \ket{\mathrm{sq}_{r+\Delta r}}$ with this unitary-transformed smeared projector.

\red{Although the projector in Eq.~\eqref{Eq: unitarysmearedpro} is expressed as an integral form, it can be implemented in practice by discretizing the parameter \(p_0\) and approximating the projector as a summation form \(\hat{P}=\sum_l p_l\hat{U}_l\). A concrete implementation algorithm of this summation-form projector is provided in Sec.~\ref{sec: Physimp}. Moreover, our numerical simulations in Sec.~\ref{sec:Simulations} confirm that sufficiently fine discretization does not introduce a noticeable error.}

We emphasize that this unitary-transformed projective squeezing scheme serves as a general framework for constructing projectors onto various resourceful subspaces in CV systems. In the following section, we demonstrate its versatility through two illustrative applications: projection onto entangled states for circuit knitting and onto non-Gaussian states for universal quantum computation.

\subsection{Applications}
Here, we introduce two types of applications. The first is projection onto the CV entangled state, i.e., the EPR and cluster states; this can realize the CV version of \textit{circuit knitting}. The circuit knitting method is a technique to decompose a large quantum circuit into smaller quantum subcircuits, and enables the simulation of quantum systems beyond the scale of the available experimental hardware. The unitary-transformed smeared projector for the EPR and cluster states can be expanded with a tensor product of displacement operators, which is a separable operation. The second application is projection onto CPS, one of the non-Gaussian states. CPS is essential for universal CV quantum computing and realizing the quantum advantages over the classical computer. We show that the smeared projector for CPS can be written as a linear combination of the product of displacement, phase-shift, and squeezing operators, which are experimentally implementable~\cite{eickbusch2022fast,sivak2023real,ayyash2024driven,del2024controlled}.

\subsubsection{Projective squeezing for EPR state and cluster state}\label{subsubsec:EPRandCluster}
An ideal EPR state is the two-mode entangled state defined as $\ket{\mathrm{EPR}}=\hat{B}\ket{x=0}\otimes\ket{p=0}$, where $\hat{B}=\exp\left(\pi(\hat{a}_1\hat{a}_2^\dagger-\hat{a}_1^\dagger\hat{a}_2)/4\right)$ is a 50:50 beam splitter. Experimentally, the EPR state is approximated with finitely squeezed vacuum states as $\ket{\mathrm{EPR}^\ast_r}=\hat{B}[\hat{S}(r)\otimes\hat{S}(-r)]\ket{0}^{\otimes2}$~\cite{ths:EPR_generation}. With the stabilizers $\hat{D}(ip_0/\sqrt{2})$ and $\hat{R}(\pi/2)\hat{D}(ix_0/\sqrt{2})\hat{R}^\dag(\pi/2)=\hat{D}(-x_0/\sqrt{2})$ of quantum states $\ket{x=0}$ and $\ket{p=0}$, the stabilizer of the EPR state can be described as
\begin{equation}
\begin{split}
    &\hat{B}\left[\hat{D}\left(i\frac{p_0}{\sqrt{2}}\right)\otimes\hat{D}\left(-\frac{x_0}{\sqrt{2}}\right)\right]\hat{B}^\dagger\\
    &=\hat{D}\left(\frac{x_0+ip_0}{\sqrt{2}}\right)\otimes\hat{D}\left(\frac{x_0-ip_0}{\sqrt{2}}\right).
    \label{eq:stabilizer_EPR}
\end{split}
\end{equation}
Given the stabilizer in Eq.~\eqref{eq:stabilizer_EPR}, the \textit{smeared} projector to the higher-squeezing-level EPR state is
\begin{align}
    \hat{P}_\mathrm{EPR}(\gamma)&=\frac{\gamma}{\pi}\int\mathrm{d}x_0\mathrm{d}p_0\exp(-\gamma(x_0^2+p_0^2))\nonumber\\
    &\hat{D}\left(\frac{x_0+ip_0}{\sqrt{2}}\right)\otimes\hat{D}\left(\frac{x_0-ip_0}{\sqrt{2}}\right)
    \label{eq:EPR_projector}
\end{align}

On the other hand, $\hat{B} \hat{x}_1 \hat{B}^\dag= (\hat{x}_1-\hat{x}_2)/\sqrt{2}$ and $\hat{B} \hat{p}_2 \hat{B}^\dag= (\hat{p}_1+\hat{p}_2)/\sqrt{2}$ combined with Eq.~\eqref{Eq: twomode} yield
\begin{equation}
\begin{aligned}
 \hat{P}_\mathrm{EPR}(\gamma)=\exp\left[-\frac{(\hat{x}_1-\hat{x}_2)^2+(\hat{p}_1+\hat{p}_2)^2}{8\gamma}\right],
\end{aligned}
\end{equation}
which approximately works as a projector for $\ket{\rm EPR} \propto \int dx \ket{x}_1 \otimes \ket{x}_2$.

Because the projection probability when increasing the squeezing level by $\Delta r$ for each mode is $\mathrm{exp}(-\Delta r)$ from Eq.~\eqref{Eq: propro} and the projection probability is invariant under unitary transformation, the projection probability for the EPR state is $\exp(-\Delta r)^2=\exp(-2\Delta r)$.

With our method, we can also project the quantum states to another type of CV entangled state, the cluster state. A CZ gate with gain $g\in\mathbb{R}$ is defined as $\hat{C}_\mathrm{z}(g)=\exp(ig\hat{x}_1\hat{x}_2)$. {In the same manner, we also define a CZ' gate as $\hat{C}_\mathrm{z}'(g)=\exp(ig\hat{p}_1\hat{p}_2)$, which we use later for circuit knitting.} Of particular note, we denote the CZ gate with gain $g=1$ as $\hat{C}_\mathrm{z}(1)=\hat{C}_\mathrm{z}$. An ideal cluster state is the two-mode entangled state defined as $\ket{\mathrm{Cluster}}=\hat{C}_\mathrm{z}\ket{p=0}^{\otimes2}$~\cite{ths:Cluster_generation}. In the same manner as the EPR state, the CZ-gate-transformed smeared projector can be considered. Because its stabilizer can be described as $\hat{C}_\mathrm{z}(g)\left[\hat{D}(x_1/\sqrt{2})\otimes\hat{D}(x_2/\sqrt{2})\right]\hat{C}_\mathrm{z}^\dagger(g)=\hat{D}((x_1+igx_2)/\sqrt{2})\otimes\hat{D}((x_2+igx_1)/\sqrt{2})$, the \textit{smeared} projector is
\begin{equation}
    \begin{split}
        \hat{P}_\mathrm{Cluster}(\gamma,g)&=\frac{\gamma}{\pi}\int\mathrm{d}x_1\mathrm{d}x_2\exp(-\gamma(x_1^2+x_2^2))\\
        &\hat{D}\left(\frac{x_1+igx_2}{\sqrt{2}}\right)\otimes\hat{D}\left(\frac{x_2+igx_1}{\sqrt{2}}\right).
        \label{eq:cluster_projector}
    \end{split}
\end{equation}

Meanwhile, $\hat{P}_{\mathrm{Cluster}}(\gamma,g)= \hat{C}_\mathrm{z}(g) [\hat{P}_{\rm asq}(\gamma) \otimes  \hat{P}_{\rm asq}(\gamma)]  \hat{C}_\mathrm{z}^\dagger(g)$ with $ \hat{C}_\mathrm{z}\hat{p}_1 \hat{C}_\mathrm{z}^\dagger(g)=\hat{p}_1- g \hat{x}_2$ and $ \hat{C}_\mathrm{z}(g)\hat{p}_2 \hat{C}_\mathrm{z}^\dagger(g)=\hat{p}_2- g \hat{x}_1$ gives
\begin{equation}\label{eq:Proj_Cluster_null}
    \hat{P}_{\mathrm{Cluster}}(\gamma,g)=\exp\left[-\frac{(\hat{p}_1-g\hat{x}_2)^2+(\hat{p}_2-g\hat{x}_1)^2}{4\gamma}\right].
\end{equation}
Thus, $\hat{P}_{\mathrm{Cluster}}(\gamma,g)$ approximately works as a projector for the state that is an eigenstate of $\hat{p}_1-g \hat{x}_2$ and $\hat{p}_2-g \hat{x}_1$ with zero-eigenvalues. Even when the input state is separable, the projected state yielded by the smeared projector $\hat{P}_{\mathrm{Cluster}}(\gamma,g)$ is an entangled state. Note that $\hat{P}_{\mathrm{Cluster}}(\gamma,g)$ approaches the exact projector for the perfect cluster state with infinite squeezing level as $\gamma$ shrinks.

Here, we discuss the projection probability achieved with the cluster-state smeared projectors. Assuming that the input state's gain and squeezing level are $g_0$ and $r_0$, respectively,  the projection probability can be calculated as 
\begin{align}
    &\bra{\mathrm{sq}_{r_0} }^{\otimes 2} \hat{C}_\mathrm{z}(g_0)^\dag \hat{P}^\dag_{\mathrm{Cluster}} (\gamma, g)\hat{P}_{\mathrm{Cluster}}(\gamma,g)\hat{C}_\mathrm{z}(g_0)\ket{\mathrm{sq}_{r_0} }^{\otimes 2}\nonumber\\
    &=\bra{\mathrm{sq}_{r_0}}^{\otimes 2} \hat{C}_\mathrm{z}(-\Delta g)\hat{P}^{\dag\otimes 2}_{\mathrm{sq}}(\gamma)\hat{P}_{\mathrm{sq}}^{\otimes 2}(\gamma)\hat{C}_\mathrm{z}(\Delta g)\ket{\mathrm{sq}_{r_0}}^{\otimes 2} ,
\end{align}
which indicates that the projection probability depends on the increased squeezing level and the gain difference $\Delta g=g_0-g$, not $g_0$ and $g$ themselves. For $\Delta g=0$, we can show that the projection probability is $\mathrm{exp}(-2 \Delta r)$ for the increased squeezing level $\Delta r$, as is the same with the EPR state.

While this method can be used for increasing the amount of entanglement, we will use this method to induce entanglement between separable subsystems in a \textit{virtual} manner, i.e., the expectation value of the entangled state can be computed by performing additional operations in the local subsystems and post-processing of the local measurement outcomes. We will show later that combining quantum gate teleportation with the virtually entangled state yields a two-mode gate operation for distant modes with only local operations and classical communications, as is detailed in Sec.~\ref{subsec:CircuitKnitting}.

\subsubsection{Projection onto the cubic phase state}\label{subsubsec:ProjectionCPS}
CPS is used to implement the cubic phase gate, one of the non-Gaussian gates~\cite{ths:CPS_generation}. Such non-Gaussian elements are essential for universal CV quantum computing and fully realizing the superiority over classical computation. An ideal CPS is defined as $\hat{U}_\mathrm{CPG}\ket{p=0}$, where $\hat{U}_\mathrm{CPG}=\exp(i\eta\hat{x}^3/3)\ (\eta\in\mathbb{R})$ is a cubic phase gate. The finite-squeezing CPS can be described as $\hat{U}_\mathrm{CPG}\ket{\mathrm{sq}_{-r}}$. Here, we call parameter $r$ the squeezing level of CPS. We show here how our projective-squeezing method can improve squeezing parameter $r$ and nonlinear parameter $\eta$. The stabilizer of CPS is $\hat{U}_\mathrm{CPG}\hat{D}(-x_0/\sqrt{2})\hat{U}^\dagger_\mathrm{CPG}=\exp(ix_0(\hat{p}-\eta\hat{x}^2))$. With Bloch-Messiah decomposition~\cite{ths:BlochMessiah}, this operator can be turned into the product of phase-shift, squeezing, another phase-shift, and displacement operators as $\hat{D}(\alpha_\mathrm{CPG}(x_0))\hat{R}(\phi_\mathrm{CPG,2}(x_0))\hat{S}(r_\mathrm{CPG}(x_0))\hat{R}(\phi_\mathrm{CPG,1}(x_0))$, where
\begin{equation}
\begin{aligned}
    \alpha_\mathrm{CPG}(x_0)&=\frac{-x_0+2i\eta x_0^2}{\sqrt{2}},\\
    \phi_\mathrm{CPG,1}(x_0)&=\arctan(\sqrt{1+\eta^2x_0^2}-\eta x_0),\\
    r_\mathrm{CPG}(x_0)&=\ln(\sqrt{1+\eta^2x_0^2}-\eta x_0),\\
    \phi_\mathrm{CPG,2}(x_0)&=-\arctan(\sqrt{1+\eta^2x_0^2}+\eta x_0).
    \label{eq:EvolutionTime}
\end{aligned}
\end{equation}
The derivation is described in Appendix~\ref{sec:CPS_decomposition}. Hence, the \textit{smeared} projector to the CPS with the parameters $(\gamma, \eta)$ $\hat{P}_\mathrm{CPS}(\gamma, \eta)= \hat{U}_{\mathrm{CPG}} \hat{P}_{\mathrm{asq}} (\gamma) \hat{U}_{\mathrm{CPG}}^\dag $ is decomposed as
\begin{equation}
\begin{aligned}
    \hat{P}&_\mathrm{CPS}(\gamma, \eta)=\sqrt{\frac{\gamma}{\pi}}\int\mathrm{d}x_0\exp(-\gamma x_0^2)\\
    &\hat{D}(\alpha_\mathrm{CPG}(x_0))\hat{R}(\phi_\mathrm{CPG,2}(x_0))\hat{S}(r_\mathrm{CPG}(x_0))\hat{R}(\phi_\mathrm{CPG,1}(x_0)).
\label{Eq: CPSpro}
\end{aligned}
\end{equation}

Meanwhile, because $\hat{U}_{\rm CPG} \hat{p} \hat{U}_{\rm CPG}^\dag = \hat{p}-\eta \hat{x}^2$, we get
\begin{equation}\label{eq:Proj_CPS_null}
\hat{P}_{\mathrm{CPS}} (\gamma, \eta) = \mathrm{exp}\bigg[-\frac{\big(\hat{p}-\eta \hat{x}^2 \big)^2}{4 \gamma}  \bigg], 
\end{equation}
which indicates that $\hat{P}_{\mathrm{CPS}} (\gamma, \eta)$ approximately works as a smeared projector for the state that is an eigenstate of $\hat{p}-\eta \hat{x}^2 $ with zero-eigenvalue, i.e., the CPS for the target nonlinear parameter $\eta$. The projection probability is given by
\begin{align}
    &\bra{\mathrm{sq}_{-r}}\hat{U}^\dag_{\mathrm{CPG}}(\eta_0)\hat{P}^\dag_{\mathrm{CPS}}(\gamma, \eta) \hat{P}_{\mathrm{CPS}}(\gamma, \eta) \hat{U}_{\mathrm{CPG}}(\eta_0) \ket{\mathrm{sq}_{-r}}\nonumber\\
    &=\bra{\mathrm{sq}_{-r}}\hat{U}^\dag_{\mathrm{CPG}}(\Delta\eta)\hat{P}^\dag_{\mathrm{sq}}(\gamma)\hat{P}_{\mathrm{sq}}(\gamma)\hat{U}_{\mathrm{CPG}}(\Delta\eta)\ket{\mathrm{sq}_{-r}},
\end{align}
which indicates that the projection probability depends on $\Delta \eta=\eta_0-\eta$ and the increased squeezing level. Notice that as $\gamma$ decreases, $\hat{P}_{\mathrm{CPS}}(\gamma, \eta)$ converges with infinite squeezing level to the exact projector for the perfect CPS. We also remark that we can project even a vacuum state onto a CPS state, dramatically easing the experimental difficulties, such as the implementation of cubic or higher order Hamiltonians to create CPS. Details are described in Appendix~\ref{sec:NoiseReduction_Projection}.


If we aim to increase the squeezing level without changing the nonlinear parameter, i.e., $\Delta\eta=0$, we get
\begin{equation}
\begin{aligned}
\hat{P}_\mathrm{CPS}(\gamma, \eta) \hat{U}_\mathrm{CPG} \ket{\mathrm{sq}_{-r}}=e^{-\Delta r/2} \hat{U}_\mathrm{CPG} \ket{\mathrm{sq}_{-(r+\Delta r)} },
\end{aligned}
\end{equation}
which indicates that we can attain a larger CPS squeezing level with projection probability $e^{-\Delta r}$.

\section{Physical implementation}
\label{sec: Physimp}
Here, we first review two implementation methods to achieve projective squeezing: LCU~\cite{ths:LCU_1, ths:LCU_2} and the VQED methods~\cite{ths:VQED}. We then describe how these methods can be applied to our unitary-transformed projective squeezing method.

\subsection{Linear combination of unitaries}
\begin{figure}
    \centering
    \includegraphics[width=1.\linewidth]{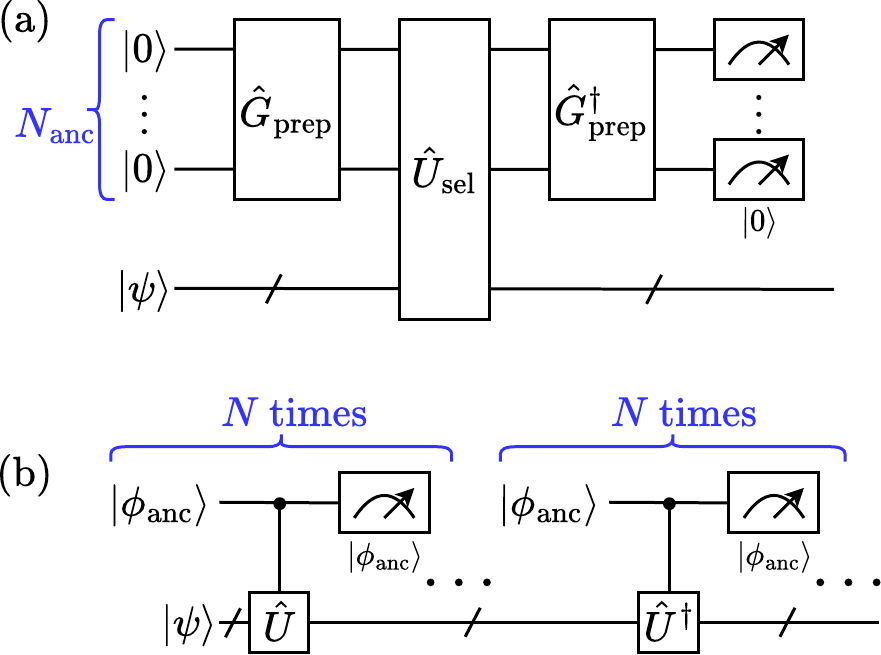}
    \caption{The circuits to implement the LCU algorithm. (a) The circuit requiring $N_{\mathrm{anc}}$ ancillary qubits. (b) The circuit requiring only one ancillary qubit.}
    \label{fig:LCU_circuits}
\end{figure}
Here, we review the LCU algorithm~\cite{ths:LCU_1, ths:LCU_2} that can be implemented with the circuits in Fig.~\ref{fig:LCU_circuits}. Let us denote the operator to be applied to the state as $\hat{P}=\sum_l p_l \hat{U}_l$ with $\hat{U}_l$ being the unitary operator and $p_l$ being the probability distribution, i.e., $p_l \geq 0$ for $\forall l$ and $\sum_l p_l =1$. Now, our target state is $\frac{\hat{P} \ket{\psi}}{\| \hat{P} \ket{\psi} \|}$ for input state $\ket{\psi}$.  In Fig.~\ref{fig:LCU_circuits}(a), we employ an ancilla state $\ket{\Phi}_{\rm anc}= \sum_l \sqrt{p_l} \ket{l}=\hat{G}_{\rm prep} \ket{0}^{\otimes N_\mathrm{anc}}$, where $\hat{G}_{\rm prep}$ is the state-preparation unitary operator and $\ket{l}$ is the orthogonal basis state. Then, we apply the select unitary operation $\hat{U}_{\rm sel}=\sum_l \ket{l}\bra{l} \otimes \hat{U}_l$ to $\ket{\Phi_{\mathrm{anc}}}\otimes\ket{\psi}$, followed by the application of $\hat{G}_{\rm prep}^\dag$ for the ancilla system. By measuring the ancilla system under a computational basis and post-selecting $\ket{0}^{\otimes N_{\rm anc}}$, we can project the state onto $\frac{\hat{P} \ket{\psi}}{\| \hat{P} \ket{\psi} \|}$ with post-selection probability $\| \hat{P} \ket{\psi} \|^2$.

Although the LCU algorithm generally requires $N_{\mathrm{anc}}$-mode entangled states as ancilla states in Fig.~\ref{fig:LCU_circuits}(a), it has been found that the LCU for the set of unitary operators $\{\hat{U}_i^n \}_{n=1}^{N} \cup \{\hat{U}_i^{\dag n} \}_{n=1}^{N}\ (N\in\mathbb{N})$ with Gaussian weight can be decomposed into the repetitive applications of LCU using only a single qubit~\cite{endo2024projective} with the circuit in Fig.~\ref{fig:LCU_circuits}(b). For the single-qubit LCU with the ancilla qubit state of $\ket{\phi_{\rm anc}}=\sqrt{p_0} \ket{0} +  \sqrt{p_1} \ket{1}$ and the select operation $\hat{U}_{\rm sel}=\ket{0}\bra{0} \otimes \hat{I} + \ket{1} \bra{1} \otimes \hat{U}$, we can project the state onto $\ket{\psi'} \propto \hat{Q} \ket{\psi}$ with $\hat{Q}=p_0 \hat{I} + p_1 \hat{U}$. Similarly, for $\hat{U}_{\rm sel}^\dag=\ket{0}\bra{0} \otimes \hat{I} + \ket{1} \bra{1} \otimes \hat{U}^\dag$, we can project the state onto $\ket{\psi''} \propto \hat{Q}^\dag \ket{\psi}$. Then, by repetitively applying the LCU for $\hat{U}_{\rm sel}$ and $\hat{U}_{\rm sel}^\dag$ and using the commutation relation $[\hat{Q}, \hat{Q}^\dag]=0$, we can, for sufficiently large $N$, project the state onto $\ket{\psi^{(N)}} \propto (\hat{Q}^\dag \hat{Q})^N \ket{\psi}= \hat{Q}^{\dag N} \hat{Q}^N \ket{\psi}$. Note that 
\begin{equation}
\begin{aligned}
\hat{Q}^N &= \sum_{k=0}^N \binom{N}{k} p_0^{N-k} p_1^k \hat{U}^k \\
& \rightarrow \sum_{k=0}^N \frac{1}{\sqrt{2 \pi N p_0 p_1}} \mathrm{exp}\bigg[- \frac{(k-N p_0)^2}{2 N p_0 p_1} \bigg] \hat{U} ^k  .
\end{aligned}
\end{equation}
 This indicates that the probability to get $\hat{U}^k$ follows a Gaussian distribution with the expectation value $N p_0 $ and the variance $N p_0 p_1$. By further applying $\hat{Q}^{\dag N}$, the expectation value of $k$ is dragged back to $0$ with the variance amplified to $2 N p_0 p_1$. Thus, we get:
\begin{equation}
\begin{aligned}
&\hat{Q}^{\dag N} \hat{Q}^N \sim \sum_{k=-N}^N \frac{1}{\sqrt{4 \pi N p_0 p_1}} \mathrm{exp}\bigg[- \frac{k^2}{4 N p_0 p_1} \bigg] \hat{U}^k,\label{eq:LCU_basic}
\end{aligned}
\end{equation}
where we define $\hat{U}^{-k}= (\hat{U} ^{\dag })^k $. 
Because the decomposition of the smeared projectors in this work involves a Gaussian distribution, this repetitive single-qubit ancilla LCU offers a hardware-efficient implementation of our method.

\subsection{Virtual quantum error detection}\label{sec:VQED}
\begin{figure}
    \centering
    \includegraphics[width=1.\linewidth]{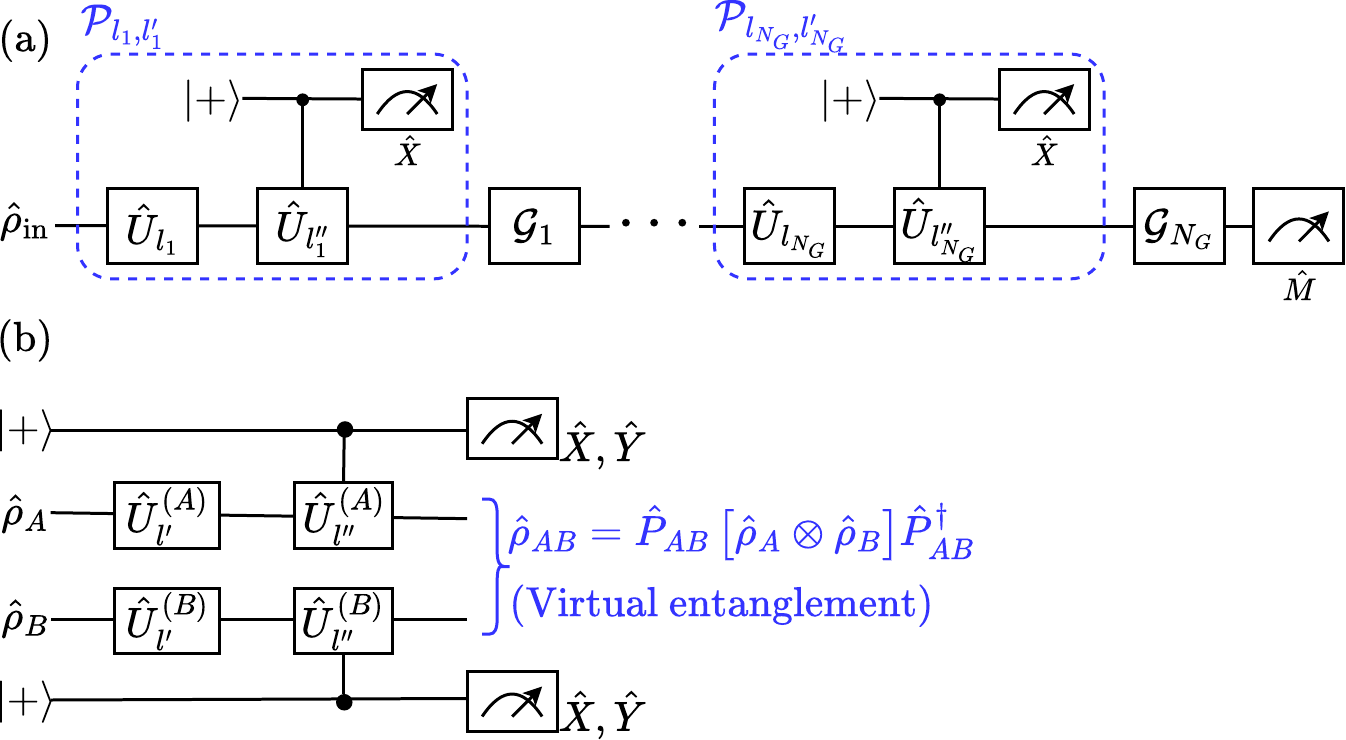}
    \caption{The circuits to implement the VQED algorithm. Here, {$\ket{+}=(\ket{0}+\ket{1})/\sqrt{2}$ is a plus state and $\hat{X}$ ($\hat{Y}$) is a Pauli-$X$ ($Y$) operator in the qubit system.} (a) The circuits to implement the ``single-mode'' VQED algorithm. By executing this circuit, we can obtain the expectation value corresponding to the state $\mathcal{G}_{N_G} \circ {\mathcal{P}}_{l_{N_G}, l'_{N_G}} \circ...\circ  \mathcal{G}_1\circ{\mathcal{P}}_{l_1, l'_1}  (\hat{\rho}_{\rm in}) $ in Eq.~\eqref{Eq:vqed}. Here, we define $\hat{U}_{l_k''}=\hat{U}_{l_k'}\hat{U}_{l_k}^\dagger$. (b) The quantum circuit for implementing the virtual projection onto the entangled subspace. Here, we denote $\hat{U}_{l''}^{(A,B)}=\hat{U}_{l'}^{(A,B)}\hat{U}_{l}^{(A,B) \dag}$. }
    \label{fig:VQED_circuits}
\end{figure}
Here, we describe the VQED method in Fig.~\ref{fig:VQED_circuits}, which allows for the computation of the expectation value of observables for a post-selected state in quantum error detection~\cite{ths:VQED}. {Compared to the LCU algorithm, the VQED algorithm offers a shallow-depth circuit with fewer controlled operations.} While the previously proposed symmetry expansion technique projects noisy quantum states onto the code space immediately before readout, the virtual quantum error detection protocol allows for the projection onto the code space during computation. 

First, we describe the quantum circuit for the VQED implementation in Fig.~\ref{fig:VQED_circuits}(a). We denote the smeared projectors as $\hat{P}=\sum_l p_l \hat{U}_l$ with $\sum_l p_l=1$ and $p_l\geq0$. We can compute the expectation value for the post-selected state as: 
\begin{equation}
\begin{aligned}\label{eq:VQED_state}
\hat{\rho}_{\rm{post}} &= \frac{\hat{\rho}_{\rm{post}}'}{\Tr[\hat{\rho}_{\rm{post}}']} \\
\hat{\rho}_{\rm{post}}' &= \mathcal{G}_{N_G} \circ \mathcal{P}_{N_G} \circ... \mathcal{G}_1 \circ \mathcal{P}_1 (\hat{\rho}_{\rm{in}}),
\end{aligned}
\end{equation}
where $\hat{\rho}_{\rm in}$ is the initial state, $N_{\rm G}$ is the number of gates, $\mathcal{G}_k$ is the $k$-th gate process, $\mathcal{P}_k (\cdot) = \hat{P}_k (\cdot) \hat{P}^\dag_k$ is the projection process before $\mathcal{G}_k$, and $\hat{P}_k = \sum_{l_k} p_{l_k} \hat{U}_{l_k} $ is the $k$-th projector. We here assume that we apply the projection each before the gate, but we can arbitrarily reduce the number of times the projection is applied.

The expectation value of an observable $\hat{M}$ with the post-selected quantum states $\langle \hat{M} \rangle _{\rm post}$ can be calculated by using Eq.~\eqref{eq:VQED_state} and substituting $\hat{P}_k = \sum_{l_k} p_{l_k} \hat{U}_{l_k} $ as: 
\begin{equation}
\begin{aligned}
&\langle \hat{M} \rangle _{\rm post} \\
&=\frac{\sum_{\vec{l}, \vec{l}'} p_{\vec{l},\vec{l}'} \Tr[\hat{M}  \mathcal{G}_{N_G} \circ {\mathcal{P}}_{l_{N_G}, l'_{N_G}} \circ...\circ \mathcal{G}_1 \circ {\mathcal{P}}_{l_1, l'_1} (\hat{\rho}_{\rm in}) ]}{\sum_{\vec{l}, \vec{l}'} p_{\vec{l},\vec{l}'} \Tr[\mathcal{G}_{N_G} \circ {\mathcal{P}}_{l_{N_G}, l'_{N_G}}  \circ...\circ \mathcal{G}_1 \circ {\mathcal{P}}_{l_1, l'_1} (\hat{\rho}_{\rm in}) ]}.
\label{Eq:vqed}
\end{aligned}
\end{equation}
Here, we denote $\mathcal{P}_{l_k, l_k'}(\cdot)=\hat{U}_{l_k} (\cdot) \hat{U}_{l'_k}^\dag$ with $p_{\vec{l},\vec{l}'}=\prod_{k=1}^{N_G} p_{l_k} p_{l'_k} $ for $\vec{l}=(l_{N_G},l_{N_G-1},...,l_1)$ and $\vec{l}'=(l'_{N_G},l'_{N_G-1},...,l'_1)$. Accordingly, we can obtain $\langle \hat{M} \rangle _{\rm post}$ by evaluating the numerator and the denominator of Eq.~\eqref{Eq:vqed} and post-processing the outcome.

\textcolor{black}{We now describe how to perform the VQED method. Because we have 
\begin{equation}
\begin{aligned}
 \sum_{l_1 l'_1} p_{l_1} p_{l'_1} \hat{U}_{l_1} \hat{\rho}_{\rm in} \hat{U}_{l'_1}^\dag= \hat{P}_1 \hat{\rho}_{\rm in} \hat{P}^\dag_1,
 \end{aligned}
\end{equation}
we can virtually project the state with the projector $\hat{P}_1$ by randomly sampling stabilizers $\{\hat{U}_{l_1} \}_{l_1}$ and $\{\hat{U}_{l'_1} \}_{l'_1}$ with the probability $p_{l_1}$ and $p_{l'_1}$, respectively. For concrete circuit implementation, owing to
\begin{equation}
\sum_{l_1 l'_1} p_{l_1} p_{l'_1} \hat{U}_{l_1} \hat{\rho}_{\rm in} \hat{U}_{l'_1}^\dag=\sum_{l_1 l'_1} p_{l_1} p_{l'_1} \hat{U}_{l_1} \hat{\rho}_{\rm in} \hat{U}_{l_1}^\dag \hat{U}_{l''_1}^\dag
\end{equation}
for $\hat{U}_{l''_1} = \hat{U}_{l'_1} \hat{U}_{l_1}^\dag  $, we can use the Hadamard test shown in Fig.~\ref{fig:VQED_circuits} (a) for computing the expectation values for the non physical states sandwiched by different unitaries~\cite{sun2022perturbative,faehrmann2022randomizing}. It has been shown that this projection can be inserted anytime during circuit execution~\cite{ths:VQED}.  For the step-by-step derivation of the VQED circuit, see Sec. I\hspace{-1.2pt}I\hspace{-1.2pt}I in Ref.~\cite{ths:VQED}.}

Then, we repeat the following process for performing the VQED method. (1) First, we generate $\vec{l}$ and $\vec{l}'$ with probability distribution $p_{\vec{l},\vec{l}'}$. \textcolor{black}{(2) Following $\vec{l}$ and $\vec{l}'$, we execute the quantum circuit in Fig.~\ref{fig:VQED_circuits}(a) for the virtual projection of the quantum states.} (3) We store the products of measurement outcomes in the computational basis $\mu_{\rm den}$ of ancilla qubits for virtual projection. Here, we write ``den'' because this term leads to the unbiased estimator of the denominator in Eq.~\eqref{Eq:vqed}. We also store outcome $m$ of the observable $\hat{M}$. Then, we calculate the products, $\mu_{\rm num} = m \mu_{\rm den}$. Here, we write ``num'' because this term leads to the unbiased estimator of the numerator in Eq.~\eqref{Eq:vqed}. By repeating this procedure, we compute the average of $\mu_{\rm num} $ and $\mu_{\rm den}$. Finally, we obtain the unbiased estimator of $\langle \hat{M} \rangle_{\rm post}$ by calculating $\frac{\langle \mu_{\rm num} \rangle}{\langle \mu_{\rm den} \rangle}$. 

Next, we describe the quantum circuit for virtually projecting the separable state onto the entangled state in the circuit knitting manner, i.e., using only local operations and post-processing, in Fig.~\ref{fig:VQED_circuits} (b) by developing the VQED method~\cite{ths:yamamoto}.  Suppose that we aim to project the separable quantum state $\hat{\rho}_A \otimes \hat{\rho}_B $ onto entangled state $\hat{\rho}_{AB}$ with the projector $\hat{P}_{AB}$ that can be linearly decomposed by separable operators, e.g., the Bell-state projector $\hat{P}_{\rm Bell}= 1/4 (\hat{I}^{(A)} \otimes \hat{I}^{(B)} + \hat{X}^{(A)} \otimes \hat{X}^{(B)}-\hat{Y}^{(A)} \otimes \hat{Y}^{(B)} +\hat{Z}^{(A)}\otimes \hat{Z}^{(B)})$ with $\hat{X},\ \hat{Y},\ \hat{Z}$ being the Pauli-$X,\ Y,\ Z$ operators. This projector can project the state onto the Bell state $\ket{\mathrm{Bell}}=1/\sqrt{2} (\ket{0}_A \ket{0}_B+ \ket{1}_A \ket{1}_B)$. Here, we assume that systems $A$ and $B$ are generally distant, so the projector can be decomposed as
\begin{equation}
\hat{P}_{AB}=\sum_{l=1}^N p_l \hat{U}_l^{(A)} \otimes \hat{U}_l^{(B)}. 
\end{equation}
For the input separable state $\hat{\rho}_A \otimes \hat{\rho}_B$, we have
\begin{equation}
\begin{aligned}
&\hat{P}_{AB} \left[\hat{\rho}_A \otimes  \hat{\rho}_B\right] \hat{P}_{AB}^\dag =\\
&\sum_{l,l'} p_{l} p_{l'} \hat{U}_{l}^{(A)} \hat{U}_{l'}^{(A) \dag} \hat{U}_{l'}^{(A)} \hat{\rho}_A \hat{U}_{l'}^{(A)\dag } 
\otimes \hat{U}_{l}^{(B)} \hat{U}_{l'}^{(B) \dag} \hat{U}_{l'}^{(B)} \hat{\rho}_B \hat{U}_{l'}^{(B) \dag}\\
&=\sum_{l,l'} p_{l} p_{l'}  \hat{U}_{l''}^{(A) \dag} \hat{U}_{l'}^{(A)} \hat{\rho}_A \hat{U}_{l'}^{(A)\dag } 
\otimes \hat{U}_{l''}^{(B) \dag} \hat{U}_{l'}^{(B)} \hat{\rho}_B \hat{U}_{l'}^{(B) \dag},
\end{aligned}
\end{equation}
with $\hat{U}_{l''}^{(A,B)}=\hat{U}_{l'}^{(A,B)}\hat{U}_{l}^{(A,B) \dag}$. Then, the quantum circuit is randomly sampled with probability $p_l p_{l'}$, see in Fig.~\ref{fig:VQED_circuits} (b), in the VQED manner, together with the measurement of the expectation values of $\hat{X}_A \otimes  \hat{X}_B$ and $\hat{Y}_A \otimes  \hat{Y}_B$. This yields 
\begin{equation}
\langle \hat{X}_A \otimes \hat{X}_B \rangle_{\rm rand} - \langle \hat{Y}_A \otimes \hat{Y}_B \rangle_{\rm rand} = \hat{P}_{AB} [\hat{\rho}_A \otimes  \hat{\rho}_B] \hat{P}_{AB}.
\end{equation}
Here, $\langle \cdot \rangle_{\rm rand}$ indicates the partial trace over the ancilla qubit with random sampling of the quantum circuit. \textcolor{black}{See Sec. S2 of Supplementary Materials in Ref.~\cite{ths:yamamoto} for derivation details. } While $\hat{P}_{AB} \left[\hat{\rho}_A \otimes  \hat{\rho}_B\right] \hat{P}_{AB}$ is an unnormalized state, we can compute the expectation value for the entangled state with the proper normalization with the projection probability as per the VQED method.


\subsection{Implementation of quantum circuit knitting}~\label{subsec:CircuitKnitting}
\begin{figure}
    \centering
    \includegraphics[width=1.\linewidth]{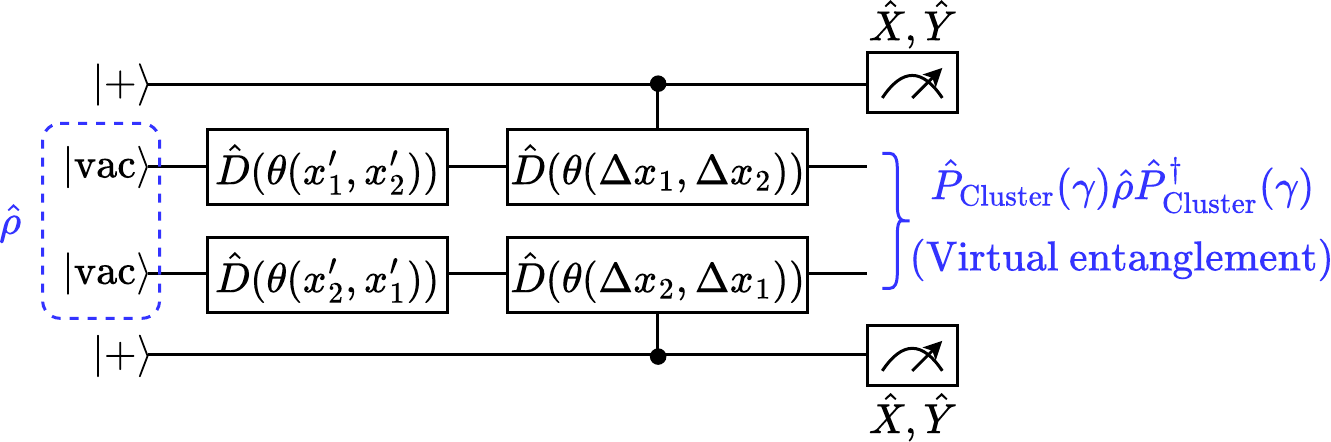}
    \caption{Quantum circuits for virtually entangling a two-mode system. Input $\ket{\mathrm{vac}}$ indicates two vacuum states, which are separable states. By iterating this circuit, we can \textit{virtually} implement the smeared projector $\hat{P}_{\mathrm{Cluster}}(\gamma, g)$. This smeared projector \textit{virtually} creates the cluster state, which is the entangled state.}
    \label{fig:Knitting_Circuits}
\end{figure}
Here, we introduce the application of our protocol to circuit knitting. Quantum circuit knitting refers to the series of techniques that allow large quantum systems to be virtually simulated from smaller ones. Here, we aim to reproduce the expectation value after virtually implementing a two-qubit gate between separable systems without any physical interaction. While quantum circuit knitting has been considered only for qubit-based quantum computing~\cite{ths:knitting_1, ths:knitting_2, ths:knitting_3, ths:knitting_4}, we propose to apply it for CV quantum computing. Our protocol relies on the fact that the smeared projector for EPR and cluster states can be decomposed into a linear combination of separable unitary operators and applied to a quantum state with LCU or VQED. However, when employing LCU, we have to prepare {controlled operations} which are entangled between the knitted quantum systems {in Fig.~\ref{fig:LCU_circuits}}. This does not lead to quantum circuit knitting because implementing such an entangling gate does not achieve the aim of quantum circuit knitting, that is, estimating the expectation value after the entangling gate \textit{without} physically realizing the gate. Therefore, we focus on the VQED implementation only needing ancillary qubits and controlled operations, which are separable between the knitted quantum systems.

Now, we can implement the CV version of the circuit knitting technique by leveraging the hybrid system of qubits and bosonic modes. To achieve CV circuit knitting, we employ the VQED method with the separable ancilla plus states $\ket{+}=(\ket{0}+\ket{1})/\sqrt{2}$ as shown in Fig.~\ref{fig:Knitting_Circuits}. In Fig.~\ref{fig:Knitting_Circuits}, we prepare a separable two-mode vacuum states and regard it as the EPR state or the cluster state with 0-dB squeezing and gain $g=0$. By \textit{virtually} increasing the squeezing level and gain through projective squeezing, shown in Fig.~\ref{fig:Knitting_Circuits}, we can \textit{virtually} entangle these two modes. More concretely, in the case of cluster-state circuit knitting, Eq.~\eqref{eq:cluster_projector} yields:
\begin{equation}\label{eq:VQED_Cluster}
\begin{aligned}
\hat{P}_{\rm Cluster}&(\gamma,g) \hat{\rho} \hat{P}^\dag_{\rm Cluster}(\gamma,g)  \\
&= \int \mathrm{d}x_1 \mathrm{d}x_2 \mathrm{d}x'_1 \mathrm{d}x'_2 p_\gamma (x_1, x_2) p_\gamma (x'_1, x'_2) \\
&\times  \hat{D}\left(\theta_g(\Delta x_1, \Delta x_2) \right) \otimes \hat{D}\left(\theta_g(\Delta x_2, \Delta x_1) \right) \\
&\times \hat{D}\left(\theta_g(x'_1, x'_2) \right)\otimes\hat{D}\left(\theta_g(x'_2, x'_1) \right)\hat{\rho}\\
&\times\hat{D}^\dag\left(\theta_g(x'_1, x'_2) \right)\otimes\hat{D}^\dag\left(\theta_g(x'_2, x'_1) \right),
\end{aligned}
\end{equation}
where $p_\gamma (a,b)= \frac{\gamma}{\pi} e^{-\gamma (a^2 + b^2)}$, $\theta_g(a, b)=(a+ig b)/2$ for $a, b\in\mathbb{R}$, $\Delta x_1 = x_1 - x'_1$ and $\Delta x_2 = x_2 - x'_2$. Therefore, because applying the smeared projector $\hat{P}_{\rm Cluster}(\gamma, g)$ to the two-mode vacuum states yields the entangled cluster states, shown in Sec.~\ref{subsubsec:EPRandCluster}, the VQED implementation with the random sampling of the quantum circuit in Fig.~\ref{fig:Knitting_Circuits} with probability $p_\gamma (x_1, x_2) p_\gamma (x'_1, x'_2)$ can simulate the expectation values for the target entangled quantum state. A similar argument holds for the EPR state.



\begin{figure}[htbp]
\centering
\includegraphics[width=1\linewidth]{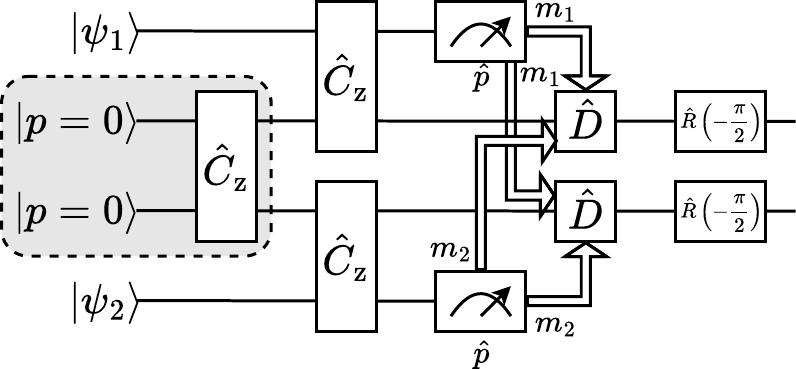}
\caption{Circuit to implement two-mode CZ' gate teleportation between inputs $\ket{\psi_1}$ and $\ket{\psi_2}$ with the ancillary cluster state, which lies in the gray area. This ancillary state can be \textit{virtually} created from a two-mode vacuum state with the VQED method in Fig.~\ref{fig:Knitting_Circuits}, which leads to circuit knitting. Here, $m_1$ and $m_2$ are the outcomes of each measurement. The mathematical formulation of this circuit is detailed in Appendix~\ref{sec:CircuitKnitting}.
}\label{fig:TwoModeGateTele}
\end{figure}

Note that our aim is to simulate a general quantum circuit with smaller quantum circuits by virtually applying two-mode operations between two distant modes, i.e., quantum circuit knitting. To achieve this, we adopt the following strategy: First, we decompose the general quantum circuit into CZ' gates and other single-mode gates. This decomposition can be achieved because only one type of two-mode gate is needed to achieve universal computing. We take the CZ' gate as part of this universal gate set. Second, we divide the large system into small subsystems and knit the subsystems with the \textit{virtually-created} CZ' gates by utilizing our projective squeezing method. To utilize our method, we show that CZ' gates can be transformed into a quantum gate teleportation circuit where a cluster state is consumed as a resource state, as is shown in Fig.~\ref{fig:TwoModeGateTele}. Then, we virtually increase the squeezing level of the cluster state from a separable two-mode vacuum state (0-dB and gain 0 cluster state) by implementing the separable displacement operation as described by Eq.~\eqref{eq:VQED_Cluster}. Implementing the procedure above allows us to simulate the CZ' gate, leading to the simulation of a larger quantum circuit. See Appendix~\ref{sec:CircuitKnitting} for the mathematical verification of the quantum teleportation circuit in Fig.~\ref{fig:TwoModeGateTele}.

To simulate a quantum circuit with smaller quantum circuits via circuit knitting, we have to measure observables more times than the situation without circuit knitting. Denoting the projection probability as $q$, we incur the sampling overhead  of $\mathcal{O}(q^{-2})$~\cite{ths:VQED}. We numerically simulate the projection  probability in Sec.~\ref{sec:Simulations} when the input state is a two-mode separable vacuum state.


\subsection{Implementation of projection onto the cubic phase state}
Here, we describe the implementations of unitary-transformed projective squeezing that use the LCU and VQED methods for cubic phase states.  

\subsubsection{Implementation with LCU}
Because the smeared projector in Eq.~(\ref{Eq: CPSpro}) is constructed from a linear combination of unitary operators with Gaussian weight, we can use the LCU implementation for projective squeezing for state preparation of CPS using just a single ancillary qubit. By substituting $\hat{U}= \mathrm{exp} (i \delta x_0 (\hat{p}-\eta \hat{x}^2) )~(\delta x_0 >0)$ for Eq.~\eqref{eq:LCU_basic}, and setting $x_0=\delta x_0 k$ and $\gamma= (4Np_0p_1 \delta x_0^2)^{-1}$, we obtain
\begin{equation}\label{eq:N_LCU}
\begin{aligned}
& \frac{1}{4 N p_0 p_1} \sum_{k=-N}^N \mathrm{exp}\bigg[-\frac{(\delta x_0 k)^2}{4 N p_0 p_1 \delta x_0 ^2} \bigg] \mathrm{exp}[i \delta x_0 k (\hat{p}-\eta \hat{x}^2) ] \\
& \rightarrow \sqrt{\frac{\gamma}{\pi}} \int \mathrm{d}x_0 \mathrm{exp}(-\gamma x_0^2) \mathrm{exp}(i x_0 (\hat{p}-\eta \hat{x}^2)) ,
\end{aligned}
\end{equation}
for $N \rightarrow \infty$. Here, we set $N=e^{2r}(e^{2\Delta r}-1)/(2p_0p_1\delta x_0^2)$ and use Eq.~\eqref{eq:VirtualSq_2}. When we apply the smeared projector in Eq.~\eqref{eq:N_LCU} to the quantum state $\hat{U}_{\mathrm{CPG}}\ket{\mathrm{sq}_r}=\exp(i\eta\hat{x}^3/3)\ket{\mathrm{sq}_r}$, we obtain $\hat{U}_{\mathrm{CPG}}\ket{\mathrm{sq}_{r+\Delta r}}$, which means that we can increase the squeezing level of the CPS while maintaining the nonlinear parameter, i.e., $\Delta \eta =0$. To implement this LCU, we need to perform controlled operations for $\mathrm{exp}(i x_0 (\hat{p}-\eta \hat{x}^2)) $. To achieve this, we decompose them into the product of controlled displacement, phase-shift, and squeezing operations following the decomposition in Eq.~(\ref{Eq: CPSpro}). Note that controlled displacement and phase-shift operations are realized in many experimental setups, and a realistic experimental implementation of controlled squeezing was recently proposed for superconducting hardware~\cite{ths:ControlSq_1,del2024controlled}. We also propose an alternative approach that realizes the controlled squeezing operation using only squeezing and controlled-phase shift gates. The details of this construction are provided in Appendix~\ref{sec:ControlledOperation}. We expect that our circuit proposal will help to reduce the experimental overhead for the state-preparation or noise suppression of the cubic phase states because it only uses Gaussian operations, aside from the experimentally-friendly controlled phase shift operations.

Note that the successful post-selection probability scales as $\exp(-\Delta r)$ for $\Delta \eta =0$, which is the same as the projection probability described in Sec.~\ref{subsubsec:ProjectionCPS}.


\subsubsection{Implementation with VQED}

VQED offers an alternative shallow-depth implementation of unitary-transformed projective squeezing that can effectively improve the state-preparation fidelity when we aim to evaluate the expectation values of observables. Here, we use Eq.~\eqref{Eq: CPSpro} and write $\hat{V}_{\mathrm{CPG}}(x_0)=\hat{D}(\alpha_\mathrm{CPG}(x_0))\hat{R}(\phi_\mathrm{CPG,2}(x_0))\hat{S}(r_\mathrm{CPG}(x_0))\hat{R}(\phi_\mathrm{CPG,1}(x_0))$. The virtual projection onto CPS can be evaluated by randomly sampling parameters $(x_0,x_0')$ following the probability distribution $\frac{\gamma}{\pi} \exp\left(-\gamma(x_0^2+x_0'^2)\right)$ and executing the circuit in Fig.~\ref{fig:VQED_circuits}(a) with $\hat{U}_{l_1}=\hat{V}_{\mathrm{CPG}}(x_0)$ and $\hat{U}_{l_1'}=\hat{V}_{\mathrm{CPG}}(x_0')$. We note that we can implement the arbitrary single-mode channel $\mathcal{G}$ after preparing the CPS. That is, after we project the input state $\hat{\rho}_{\mathrm{in}}$ with the smeared projector $\hat{P}_{\mathrm{CPS}}(\gamma, \eta)$ as $\mathcal{P}_{\mathrm{CPS}}(\hat{\rho}_{\mathrm{in}})=\hat{P}_{\mathrm{CPS}}(\gamma, \eta)\hat{\rho}_{\mathrm{in}}\hat{P}_{\mathrm{CPS}}^\dagger(\gamma, \eta)$, we can implement the single-mode channel $\mathcal{G}$ as $\mathcal{G}\circ\mathcal{P}_{\mathrm{CPS}}(\hat{\rho}_{\mathrm{in}})$, as is shown in Fig.~\ref{fig:VQED_circuits}(a). For projection probability $q$, the sampling overhead for this method scales with $\mathcal{O}(q^{-2})$~\cite{ths:VQED}. For $\Delta \eta=0$, the sampling overhead reads $\mathcal{O}(e^{2 \Delta r})$.

\section{Numerical simulations}\label{sec:Simulations}

\begin{figure}[t]
\centering
\includegraphics[width=1.\linewidth]{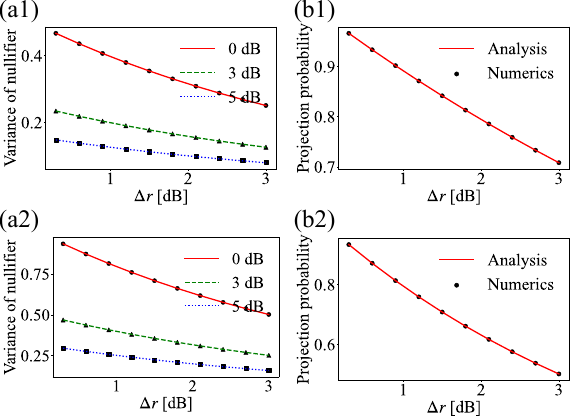}
\caption{Comparison of numerical simulations and analytical results for the projection of CPS and the cluster state with $\Delta \eta=0$ and $\Delta g=0$. (a1) The variance of the CPS nullifier. (a2) The variance of the cluster-state nullifier. (b1) The projection probability of CPS. (b2) The projection probability of the cluster state. In these plots, the markers represent the numerical results and the lines represent the analytical results. Details are described in the main text.}\label{fig:plot}
\end{figure}

\begin{figure}[t]
\centering
\includegraphics[width=1.0\linewidth]{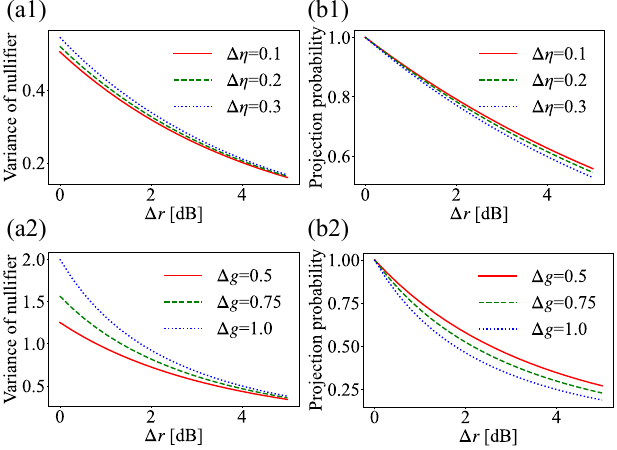}
\caption{Numerical simulations about the effect of increasing nonlinear parameter $\eta$ for CPS and gain $g$ for the cluster state. Here, we plot their results by increasing the parameter $\eta$ by 0.1, 0.2, and 0.3 for CPS and the parameter $g$ by 0.5, 0.75, and 1.0 for the cluster state. (a1) The variance of the CPS nullifier. (a2) The variance of the cluster-state nullifier. (b1) The projection probability for CPS. (b2) The projection probability for the cluster state.}\label{fig:plot_eta}
\end{figure}

\begin{figure}[t]
\centering
\includegraphics[width=1.0\linewidth]{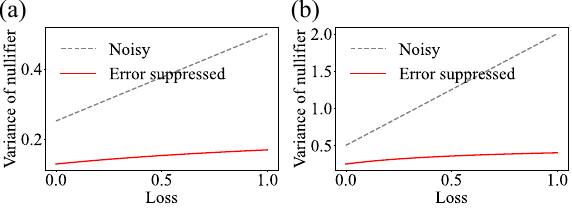}
\caption{Numerical simulations addressing the effect of photon loss. Here, we consider the CPS and the cluster state after the photon-loss process and implement the smeared projector for such noisy states. Then, we compare the results before and after the projections. (a) The CPS nullifier before (Noisy) and after (Error suppressed) projection. (b) The cluster state nullifier before (Noisy) and after (Error suppressed) projection.}\label{fig:plot_loss}
\end{figure}

\begin{figure}[t]
\centering
\includegraphics[width=1.0\linewidth]{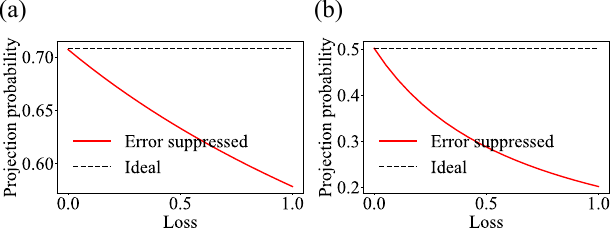}
\caption{\red{Numerical simulations of projection probability. All of the conditions, other than the vertical axis, are the same as Fig.~\ref{fig:plot_loss}. (a) The projection probability of CPS after projection (Error suppressed). (b) The projection probability of the cluster state after projection (Error suppressed).}}\label{fig:plot_projectionprobability}
\end{figure}

This section describes the results of numerical simulations for unitary-transformed projective squeezing for CPS and cluster states. We numerically evaluate the variance of nullifiers and the projection probability. We also confirm that our protocol can mitigate the effect of photon loss due to the projection onto the target state.

We simulate the application of the smeared projectors in Eq.~\eqref{eq:cluster_projector} and Eq.~\eqref{Eq: CPSpro}. Here, parameter $\gamma$ depends on the projectively increased squeezing level $\Delta r$, as is shown in Eq.~\eqref{eq:VirtualSq_2}. As for the parameters $x_0,\ x_1$ and $x_2$ in Eq.~\eqref{Eq: CPSpro} or Eq.~\eqref{eq:cluster_projector}, to avoid the integration, we divide the range $[-2\sqrt{\gamma/2},2\sqrt{\gamma/2}]$ into $30$ steps, and consider the linear combination of the displacement operator with the $30$ points of $x_0,\ x_1$ and $x_2$. Such discretization of the parameters is needed to implement VQED and LCU in actual experiments; however, later numerical simulations confirm that such discretization does not cause the error, and the results of numerical simulations agree well with the analytical results.

First, we evaluate the variance of the CPS nullifier $\expval{[\Delta(\hat{p}-\eta\hat{x}^2)]^2}$ in Fig.~\ref{fig:plot}(a1) and the cluster-state nullifier $\expval{[\Delta(\hat{p}_1-\hat{x}_2)]^2}+\expval{[\Delta(\hat{p}_2-\hat{x}_1)]^2}$ in Fig.~\ref{fig:plot}(a2) for $\Delta \eta= \Delta g=0$ to compare the numerical results with the analytical ones. Here, $\expval{[\Delta(\cdot)]^2}$ denotes the variance. In Fig.~\ref{fig:plot}(a1) and (a2), the vertical axis represents these nullifiers. The horizontal axis represents the projectively increased squeezing level $\Delta r$. We increase the squeezing level from the initial states whose squeezing levels $r$ are $0, 3, 5$ dB, which is shown by the red, green, and blue lines, respectively. The lines represent the analytical results, and the markers represent the numerical results. The analytical results plot the value $\exp(-2(r+\Delta r))/2$ for the single-mode CPS and $\exp(-2(r+\Delta r))$ for the two-mode cluster state. The derivations of these analytical variances of nullifiers are described in Appendix~\ref{sec:Nullifiers}. Figures~\ref{fig:plot}  (a1) and (a2) show that the variance of the nullifier falls as the squeezing level rises for CPS and the cluster state.

We also simulate the projection probability. Figure~\ref{fig:plot}(b1) presents the CPS result, and Fig.~\ref{fig:plot}(b2) presents the cluster state result. The horizontal axis plots the projectively increased squeezing level $\Delta r$, while the vertical axis presents the projection probability. Here, we set the initial squeezing level as 0 dB but note that the projection probability depends on just the projectively increased squeezing level $\Delta r$, not on the other parameters such as the initial squeezing level. The lines plot the analytical results, and the markers represent the numerical results. The analytical results plot the value $\exp(-\Delta r)$ for single-mode CPS and $\exp(-2\Delta r)$ for the two-mode cluster state. The numerical results plot the value $\bra{\psi}\hat{P^\dagger}\hat{P}\ket{\psi}$ for each input state $\ket{\psi}$ and each smeared projector $\hat{P}$. The numerical results of the projection probability agree with the analytical results. This agreement indicates that discretization of the parameters such as $x_0,\ x_1,$ and $x_2$ does not cause errors.\\

Second, we numerically simulate the cases of $\Delta \eta \neq 0$ and  $\Delta g \neq 0$, i.e., increasing the gain and the nonlinear parameter. The numerical results are shown in Fig.~\ref{fig:plot_eta}. In these plots, we prepare the vacuum state ($r=0$, $\eta=0$, and $g=0$) as the initial state. Then, we aim to increase the squeezing level up to 5 dB. At the same time, we increase the nonlinear parameter $\eta$ by 0.1, 0.2, and 0.3 for CPS and the gain $g$ by 0.5, 0.75, and 1.0 for the cluster state. Figure~\ref{fig:plot_eta}(a1) ((a2)) evaluates the variance of the CPS nullifier (the cluster state). Even though the variances increase with $\Delta \eta$ ($\Delta g$), the variances of the quantum state nullifiers after the projections are suppressed as we increase the squeezing parameter. This result indicates that our unitary-transformed projective squeezing method allows the quantum noise to be squeezed even when this increases the nonlinear parameter $\eta$ (gain $g$). Figure~\ref{fig:plot_eta}(b1) ((b2)) evaluates the projection probability of CPS (the cluster state). We also find that the projection probability decreases as $\Delta \eta$ ($\Delta g$) increases.

Finally, we also numerically confirm that our method can mitigate the effect of photon loss. Figure~\ref{fig:plot_loss}(a) presents the CPS result while Fig.~\ref{fig:plot_loss}(b) presents that of the cluster state.  Here, we simulated the photon-loss dynamics described by the completely positive trace-preserving (CPTP) map described below
\begin{align}\label{eq:LossChannel}
    \mathcal{N}(L)\{\hat{\rho}\}&=\sum_{n=0}^\infty \hat{E}_n(L)\hat{\rho}\hat{E}_n(L)^\dagger,\nonumber\\ 
    \hat{E}_n(L)&=\left(\frac{L}{1-L}\right)^{\frac{n}{2}}\frac{\hat{a}^n}{\sqrt{n!}}(\sqrt{1-L})^{\hat{a}^\dag\hat{a}},
\end{align}
where $L$ represents a loss rate~\cite{ths:LossChannel}. This formulation is used for the single-mode state, but we can directly apply it to the two-mode state by considering the photon loss of each mode. 

Figures~\ref{fig:plot_loss}(a) and (b) show the results for CPS and the cluster state. In this simulation, we first prepare the 3-dB cubic phase (cluster) state and simulate loss $L$, following Eq.~\eqref{eq:LossChannel}. Then, we projectively squeeze the noisy state by 3 dB and measure the variance of the nullifier of the cubic phase (cluster) state. This result is labeled ``Error suppressed'' in Fig.~\ref{fig:plot_loss}(a) and (b). We also evaluate the variance of the nullifier of the cubic phase (cluster) state without projective squeezing. This result is labeled ``Noisy'' in Fig.~\ref{fig:plot_loss}(a) and (b). The results indicate that we can robustly project the noisy and mixed state to the ideal state, and so mitigate the effect of photon loss. \red{We also simulate the projection probability under photon-loss dynamics in Fig.~\ref{fig:plot_projectionprobability}. Compared with Fig.~\ref{fig:plot_loss}, only the vertical axis has changed from nullifier variance to projection probability; other conditions are the same. The label ‘Error suppressed’ denotes the numerical projection probability after photon loss, and ‘Ideal’ denotes the analytical value without photon loss. Although the projection probability decreases with the photon-loss rate, its order of magnitude remains stable, indicating that photon loss does not drastically impair our scheme.}

\section{Discussions and Conclusions}
In this paper, we proposed methods to project CV quantum states onto states that are unitary-transformed from a squeezed vacuum by constructing a smeared projector for the target subspace. To implement projective squeezing, we considered the LCU method and the VQED method. With the LCU method, we can physically project the state onto EPR and cluster states with higher squeezing levels and gain and CPS with higher squeezing levels and nonlinearity. The LCU method requires the iteration of controlled-unitary operation and post-selection; see Fig.~\ref{fig:LCU_circuits}. Meanwhile, the VQED method can be implemented with the relatively simpler circuits of Fig.~\ref{fig:VQED_circuits} compared to the LCU circuits, but we can obtain only the expectation value. We utilize this VQED method and quantum two-mode gate teleportation to establish the CV version of the circuit knitting technique, which allows small devices to realize large-scale quantum computing. 

We also numerically verified our method. We first compared the numerical and analytical results of the variance of the nullifiers and the projection probability for the cluster and CPS under $\Delta g=0$ and $\Delta \eta=0$. The variance of the nullifiers can be used to measure the quality of the projectively squeezed state. The projection probability can be used to estimate the costs incurred when implementing the projective squeezing method. For $\Delta \eta \neq 0$ and  $\Delta g \neq 0$, we also confirmed that the state approaches the target state with smaller projection probabilities for larger $\Delta g (\eta)$.  We also numerically confirmed that the variance of nullifiers of the state affected by the photon loss can be suppressed by our method.

We discuss here some future directions for our proposal. First, we may be able to utilize our method on platforms other than superconducting hardware. In this paper, we considered superconducting hardware because it offers the experimentally implementable controlled unitary operations required for both the LCU and the VQED methods. However, such controlled unitary operations can also be realized in other systems, such as cavity QED systems~\cite{hacker2019deterministic} and ion-trap systems~\cite{de2022error}. \red{Second, we do not directly simulate the LCU and VQED methods, and hence, realistic experimental conditions are not fully captured. While we approximate such effects via photon-loss dynamics during gate operation (using parameters drawn from conventional platforms; see Appendix~\ref{sec:NoiseReduction_Projection}), a more detailed and method-specific modeling is an important direction for future work.} \red{In addition}, while we proposed two practical applications of unitary-transformed projective squeezing (circuit knitting and the preparation of a non-Gaussian state), searching for other practical applications is interesting. \red{Moreover}, although we decompose the smeared projector for the target manifold with $\{\hat{U} \hat{D}(\alpha) \hat{U}^\dag\}_\alpha$, there may be other basis choices for efficiently decomposing the target smeared projector with reduced sampling overhead and hardware requirements. Finally, because the VQED method can be considered a subclass of quantum error mitigation (QEM) ~\cite{endo2021hybrid,cai2023quantum} and can be combined with QEM strategies, searching for efficient combinations of VQED-based projective squeezing with QEM methods is a question of practical importance.

\section*{Acknowledgments}
 This project is supported by Moonshot R\&D, JST, Grant No.\,JPMJMS2061; MEXT Q-LEAP Grant No.\,JPMXS0120319794, and No.\, JPMXS0118068682 and PRESTO, JST, Grant No.\, JPMJPR2114,  No. \,JPMJPR1916, JST CREST Grant No. JPMJCR23I4, and JST FOREST Grant No. \,JPMJFR223R, Japan. We acknowledge useful discussions with Kaoru Yamamoto.


%




\appendix

\section{Derivation of projective squeezing for squeezed vacuum}\label{sec:ProjSq_SqVac}

Here, we review the projective squeezing for the squeezed-vacuum state discussed in Ref~\cite{endo2024projective}. The squeezed vacuum state with squeezing level $r$ can be written in $x$ basis as
\begin{equation}\label{eq:x-sqvac}
    \ket{\mathrm{sq}_r}=\pi^{-\frac{1}{4}}e^{\frac{r}{2}}\int\mathrm{d}x\exp\left(-\frac{e^{2r}}{2}x^2\right)\ket{x}.
\end{equation}
By applying the smeared projector $\hat{P}_\mathrm{sq}(\gamma)=\int\mathrm{d}p_0\sqrt{\gamma/\pi}\exp(-\gamma p_0^2)\hat{D}(ip_0/\sqrt{2})$ to Eq.~\eqref{eq:x-sqvac}, the projectively squeezed state can be written as
\begin{equation}
    \begin{split}
        \hat{P}_\mathrm{sq}(\gamma)\ket{\mathrm{sq}_r}&=\sqrt{\frac{\gamma}{\pi}}\pi^{-\frac{1}{4}}e^{\frac{r}{2}}\int\mathrm{d}x\mathrm{d}p_0\exp(-\gamma p_0^2)\exp(ip_0x)\\
        &\exp\left(-\frac{e^{2r}}{2}x^2\right)\ket{x}\\
        &=\pi^{-\frac{1}{4}}e^{\frac{r}{2}}\int\mathrm{d}x\exp\left(-\frac{x^2}{4\gamma}-\frac{e^{2r}}{2}x^2\right)\ket{x}.
    \end{split}
\end{equation}
Here, we can define the projectively increased squeezing level $\Delta r$ as
\begin{equation}
    \frac{1}{4\gamma}+\frac{e^{2r}}{2}=\frac{e^{2(r+\Delta r)}}{2}.
\end{equation}
With the parameter $\Delta r$, the projectively squeezed vacuum can be written as
\begin{align}
    \hat{P}&_\mathrm{sq}(\gamma)\ket{\mathrm{sq}_r}\nonumber\\
        &=e^{-\frac{\Delta r}{2}}\left[\pi^{-\frac{1}{4}}e^{\frac{(r+\Delta r)}{2}}\int\mathrm{d}x\exp\left(-\frac{e^{2(r+\Delta r)}}{2}x^2\right)\ket{x}\right]\nonumber\\
        &=e^{-\frac{\Delta r}{2}}\ket{\mathrm{sq}_{r+\Delta r}}.
\end{align}
Hence, we can projectively increase the squeezing level $r$ by $\Delta r$, with projection probability $\exp({-\Delta r/2})^2=\exp(-\Delta r)$.

\section{Derivation of projective squeezing for cubic phase state}\label{sec:CPS_decomposition}

Here, we describe the derivation of projective squeezing for CPS. First, the CPS stabilizer $\hat{U}_\mathrm{CPG}\hat{D}(-x_0/\sqrt{2})\hat{U}^\dagger_\mathrm{CPG}$ can be derived as
\begin{equation}\label{eq:stablizier_cpg}
    \begin{split}
        \hat{U}&_\mathrm{CPG}\hat{D}\left(-\frac{x_0}{\sqrt{2}}\right)\hat{U}^\dagger_\mathrm{CPG}\\
        &=\exp\left(i\frac{\eta}{3}\hat{x}^3\right)\exp\left(-ix_0\hat{p}\right)\exp\left(-i\frac{\eta}{3}\hat{x}^3\right)\\
        &=\exp\left(-ix_0\left(\exp\left(i\frac{\eta}{3}\hat{x}^3\right)\hat{p}\exp\left(-i\frac{\eta}{3}\hat{x}^3\right)\right)\right)\\
        &=\exp\left(-ix_0\left(\hat{p}-\eta\hat{x}^2\right)\right),
    \end{split}
\end{equation}
with the Campbell-Baker-Hausdorff formula
\begin{equation}
    e^{\hat{A}}\hat{B}e^{-\hat{A}}=\hat{B}+[\hat{A},\hat{B}]+\frac{1}{2!}\left[\hat{A},[\hat{A},\hat{B}]\right]+\cdots.
\end{equation}
The stabilizer in Eq.~\eqref{eq:stablizier_cpg} can be decomposed into a product of the squeezing and phase-shift operators with the Bloch-Messiah decomposition because the stabilizer is a Gaussian operator~\cite{ths:BlochMessiah}. To decompose the stabilizer, first, we describe the action of the operator in Eq.~\eqref{eq:stablizier_cpg} in the Heisenberg picture as
\begin{equation}
    \begin{split}
        \exp&\left(-ix_0\left(\hat{p}-\eta\hat{x}^2\right)\right)\begin{pmatrix}
            \hat{x}\\ \hat{p}
        \end{pmatrix}\exp\left(ix_0\left(\hat{p}-\eta\hat{x}^2\right)\right)\\
        &=\begin{pmatrix}
            1&0\\-2\eta x_0&1
        \end{pmatrix}
        \begin{pmatrix}
            \hat{x}\\ \hat{p}
        \end{pmatrix}+
        \begin{pmatrix}
            -x_0\\2\eta x_0^2
        \end{pmatrix},
    \end{split}
\end{equation}
with the Campbell-Baker-Hausdorff formula. Then, the coefficient matrix is decomposed into
\begin{align}
    &\begin{pmatrix}
            1&0\\-2\eta x_0&1
    \end{pmatrix}=
    \begin{pmatrix}
        \cos\phi_\mathrm{CPG,2}(x_0)&-\sin\phi_\mathrm{CPG,2}(x_0)\\ \sin\phi_\mathrm{CPG,2}(x_0)&\cos\phi_\mathrm{CPG,2}(x_0)
    \end{pmatrix}\nonumber\\
    &\times
    \begin{pmatrix}
        e^{-r_\mathrm{CPG}(x_0)}&0\\0&e^{r_\mathrm{CPG}(x_0)}
    \end{pmatrix}\nonumber\\
    &\times
    \begin{pmatrix}
        \cos\phi_\mathrm{CPG,1}(x_0)&-\sin\phi_\mathrm{CPG,1}(x_0)\\ \sin\phi_\mathrm{CPG,1}(x_0)&\cos\phi_\mathrm{CPG,1}(x_0)
    \end{pmatrix},
\end{align}
where $\phi_\mathrm{CPG,1}(x_0)=\arctan(\sqrt{1+\eta^2x_0^2}-\eta x_0)$, $r_\mathrm{CPG}(x_0)=\ln(\sqrt{1+\eta^2x_0^2}-\eta x_0)$, and $\phi_\mathrm{CPG,2}(x_0)=-\arctan(\sqrt{1+\eta^2x_0^2}+\eta x_0)$. Hence, we can implement the operation $\exp\left(ix_0\left(\hat{p}-\eta\hat{x}^2\right)\right)$ with the phase-shift, squeezing, another phase-shift, and displacement operations, as is described in the main text.

\section{CZ'-gate-teleportation circuit}\label{sec:CircuitKnitting}

\begin{figure*}[htbp]
\centering
\includegraphics[width=1\linewidth]{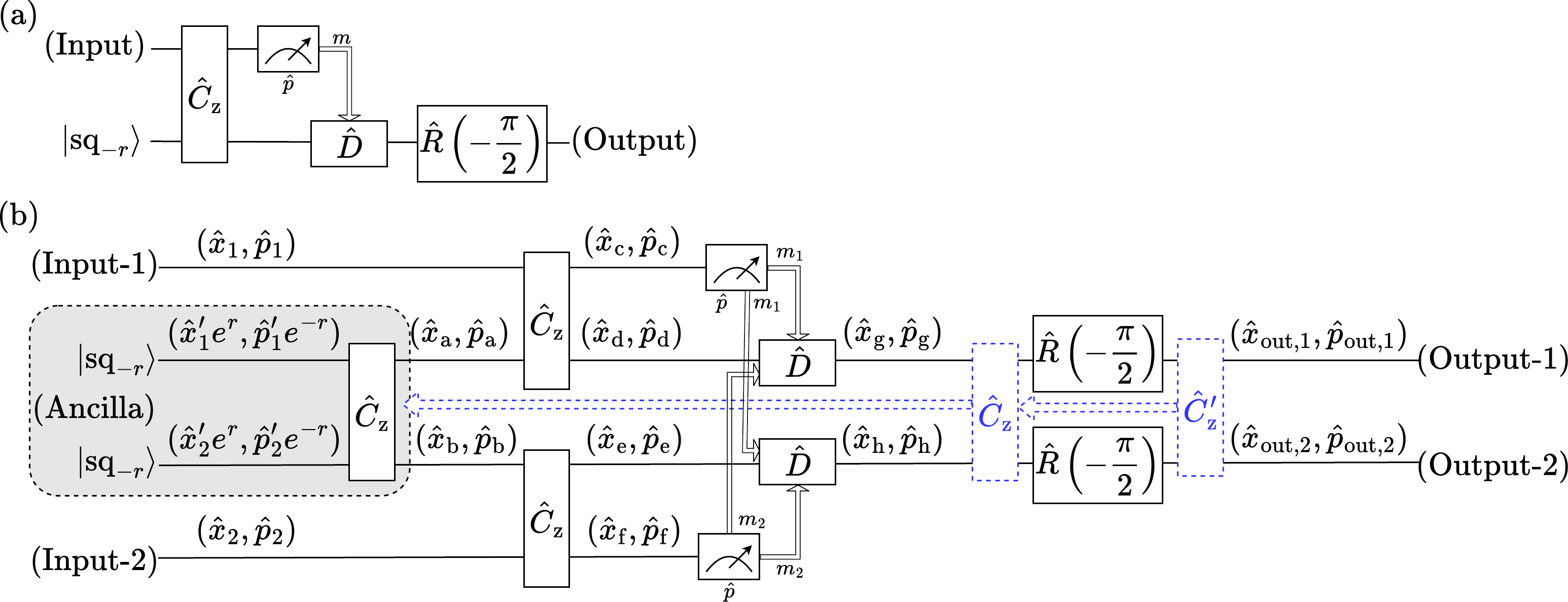}
\caption{Circuits to implement two-mode CZ' gate teleportation with ancillary cluster state. (a) The quantum teleportation circuit whose output is the same as the input. This circuit is used to construct the circuit in Fig.~\ref{fig:TwoModeGateTele_ap}(b). (b) The quantum gate-teleportation circuit. With this circuit, we implement the CZ' gate between Input-1 and Input-2 by consuming the ancillary cluster state shown in the gray area. The mathematical formulation of this circuit is described in Appendix~\ref{sec:CircuitKnitting}.
}\label{fig:TwoModeGateTele_ap}
\end{figure*}

Here, we derive the mathematical verification of CZ'-gate teleportation in Fig.~\ref{fig:TwoModeGateTele} of the main text by using Fig.~\ref{fig:TwoModeGateTele_ap}. First, we explain how to obtain the quantum circuit for the teleportation-based two-mode operation. Figure~~\ref{fig:TwoModeGateTele_ap}(a) is a single-mode quantum teleportation circuit whose output is the same as the input~\cite{ths:HalfTele}. Then, as is shown in Fig.~\ref{fig:TwoModeGateTele_ap}(b), we prepare two sets of the quantum teleportation circuit in Fig.~\ref{fig:TwoModeGateTele_ap}(a) and we directly implement the CZ' gate between Output-1 and Output-2. Then, by moving the CZ' gate to the left side of the circuit in Fig.~\ref{fig:TwoModeGateTele_ap}(b), we finally reach a gate-teleportation circuit that consumes the two-mode cluster state as the ancillary state. We note that the EPR state cannot be used directly for the strategy above because the beam splitter operation used to create the EPR state from squeezed vacuum states does not commute with the CZ gate and thus the beam splitter operation cannot move to the left side of the circuit in Fig.~\ref{fig:TwoModeGateTele_ap}(b).

Next, we mathematically verify that the circuit in Fig.~\ref{fig:TwoModeGateTele_ap} (b) can perform teleportation-based two-mode operation. We denote the quadrature amplitudes of the two-mode input state as $(\hat{x}_1,\hat{p}_1)$ and $(\hat{x}_2,\hat{p}_2)$. We also denote the ancillary two-mode $p$-squeezed states with squeezing level $r$ as $(\hat{x}_1'e^r,\hat{p}_1'e^{-r})$ and $(\hat{x}_2'e^r,\hat{p}_2'e^{-r})$. Here, $(\hat{x}_i', \hat{p}_i')\ (i=1,2)$ represents the quadrature of the vacuum state. After the CZ gate between the ancillary $p$-squeezed states, the quadrature amplitudes of these modes are written in Heisenberg picture as
\begin{align}
     \begin{pmatrix}
        \hat{x}_\mathrm{a}\\ \hat{p}_\mathrm{a}
    \end{pmatrix}
    &=
    \begin{pmatrix}
        \hat{x}_1'e^r\\ \hat{p}_1'e^{-r}+x_2'e^r
    \end{pmatrix}
    ,\ \\
    \begin{pmatrix}
        \hat{x}_\mathrm{b}\\ \hat{p}_\mathrm{b}
    \end{pmatrix}
    &=
    \begin{pmatrix}
        \hat{x}_2'e^r\\ \hat{p}_2'e^{-r}+x_1'e^r
    \end{pmatrix}.
\end{align}
Then, we implement the CZ gate between input modes and ancillary modes. The quadrature amplitudes are changed as
\begin{align}
        \begin{pmatrix}
        \hat{x}_\mathrm{c}\\ \hat{p}_\mathrm{c}
    \end{pmatrix}
    &=
    \begin{pmatrix}
        \hat{x}_1 \\ \hat{p}_1+\hat{x}_1'e^r
    \end{pmatrix}
    ,\ \\
    \begin{pmatrix}
        \hat{x}_\mathrm{d}\\ \hat{p}_\mathrm{d}
    \end{pmatrix}
    &=
    \begin{pmatrix}
        \hat{x}_1'e^r\\ \hat{p}_1'e^{-r}+x_2'e^r+\hat{x}_1
    \end{pmatrix},\ \\
    \begin{pmatrix}
        \hat{x}_\mathrm{e}\\ \hat{p}_\mathrm{e}
    \end{pmatrix}
    &=
    \begin{pmatrix}
        \hat{x}_2'e^r\\ \hat{p}_2'e^{-r}+x_1'e^r+\hat{x}_2
    \end{pmatrix}
    ,\ \\
    \begin{pmatrix}
        \hat{x}_\mathrm{f}\\ \hat{p}_\mathrm{f}
    \end{pmatrix}
    &=
    \begin{pmatrix}
        \hat{x}_2 \\ \hat{p}_2+\hat{x}_2'e^r
    \end{pmatrix}.
\end{align}
We measure $(\hat{p}_\mathrm{c},\hat{p}_\mathrm{f})$, and denote the measurement result as $(m_1, m_2)$, respectively. Then, we implement the feedforward displacement operation $\hat{D}(-(m_1+im_2)/2)$ and $\hat{D}(-(m_2+im_1)/2)$ to quadrature $(\hat{x}_\mathrm{d},\hat{p}_\mathrm{d})$ and $(\hat{x}_\mathrm{e},\hat{p}_\mathrm{e})$. After this feedforward operation, these quadrature amplitudes can be written as
\begin{align}
    \begin{pmatrix}
        \hat{x}_\mathrm{g}\\ \hat{p}_\mathrm{g}
    \end{pmatrix}
    &=
    \begin{pmatrix}
        -\hat{p}_1\\ \hat{x}_1-\hat{p}_2+\hat{p}_1'e^{-r}
    \end{pmatrix}
    ,\ \\
    \begin{pmatrix}
        \hat{x}_\mathrm{h}\\ \hat{p}_\mathrm{h}
    \end{pmatrix}
    &=
    \begin{pmatrix}
        -\hat{p}_2\\ \hat{x}_2 -\hat{p}_1+\hat{p}_2'e^{-r}
    \end{pmatrix}.
\end{align}
After the phase shift of $-\pi/2$, the output is described as
\begin{align}
    \begin{pmatrix}
        \hat{x}_\mathrm{out,1}\\ \hat{p}_\mathrm{out,1}
    \end{pmatrix}
    &=
    \begin{pmatrix}
        \hat{x}_1-\hat{p}_2+\hat{p}_1'e^{-r}\\ \hat{p}_1
    \end{pmatrix}
    ,\ \\
    \begin{pmatrix}
        \hat{x}_\mathrm{out,2}\\ \hat{p}_\mathrm{out,2}
    \end{pmatrix}
    &=
    \begin{pmatrix}
        \hat{x}_2 -\hat{p}_1+\hat{p}_2'e^{-r}\\ \hat{p}_2
    \end{pmatrix},
\end{align}
which coincides with the CZ' gate operation with gain $g=1$ at the limit of infinite squeezing $r\rightarrow\infty$.\par

\section{Derivations of variance of nullifiers}\label{sec:Nullifiers}

Here, we derive the variance of the nullifiers for the $x$-squeezed vacuum state, CPS, and the cluster state.

First, the variance of the nullifier $\expval{[\Delta(\hat{x})])^2}$ for the $x$-squeezed vacuum state $\ket{\mathrm{sq}_{r}}$ can be calculated as
\begin{align}
    &\expval{[\Delta(\hat{x})])^2}\nonumber\\
    =&\bra{\mathrm{sq}_{r}}\hat{x}^2\ket{\mathrm{sq}_{r}}-\bra{\mathrm{sq}_{r}}\hat{x}^2\ket{\mathrm{sq}_{r}}^2\nonumber\\
    =&\bra{0}\hat{S}^\dagger(r)\hat{x}^2\hat{S}(r)\ket{0}-\bra{0}\hat{S}^\dagger(r)\hat{x}\hat{S}(r)\ket{0}^2\nonumber\\
    =&\bra{0}(\hat{x}e^{-r})^2\ket{0}-\bra{0}\hat{x}e^{-r}\ket{0}^2\nonumber\\
    =&\frac{e^{-2r}}{2}-0\nonumber\\
    =&\frac{e^{-2r}}{2}.
\end{align}

Second, the variance of the nullifier $\expval{[\Delta(\hat{p}-\eta\hat{x}^2)]^2}$ for CPS $\hat{U}_{\mathrm{CPG}}\ket{\mathrm{sq}_{-r}}$ can be calculated using
\begin{align}
    &\expval{[\Delta(\hat{p}-\eta\hat{x}^2)]^2}\nonumber\\
    =&\bra{\mathrm{sq}_{-r}}\hat{U}_{\mathrm{CPG}}^\dagger(\hat{p}-\eta\hat{x}^2)^2\hat{U}_{\mathrm{CPG}}\ket{\mathrm{sq}_{-r}}\nonumber\\
    &-\bra{\mathrm{sq}_{-r}}\hat{U}_{\mathrm{CPG}}^\dagger(\hat{p}-\eta\hat{x}^2)\hat{U}_{\mathrm{CPG}}\ket{\mathrm{sq}_{-r}}^2\nonumber\\
    =&\bra{\mathrm{sq}_{-r}}\hat{U}_{\mathrm{CPG}}^\dagger(\hat{U}_{\mathrm{CPG}}\hat{p}\hat{U}_{\mathrm{CPG}}^\dagger)^2\hat{U}_{\mathrm{CPG}}\ket{\mathrm{sq}_{-r}}\nonumber\\
    &-\bra{\mathrm{sq}_{-r}}\hat{U}_{\mathrm{CPG}}^\dagger(\hat{U}_{\mathrm{CPG}}\hat{p}\hat{U}_{\mathrm{CPG}}^\dagger)\hat{U}_{\mathrm{CPG}}\ket{\mathrm{sq}_{-r}}^2\nonumber\\
    =&\bra{\mathrm{sq}_{-r}}\hat{p}^2\ket{\mathrm{sq}_{-r}}-\bra{\mathrm{sq}_{-r}}\hat{p}\ket{\mathrm{sq}_{-r}}^2\nonumber\\
    =&\bra{0}\hat{S}^\dagger(-r)\hat{p}^2\hat{S}(-r)\ket{0}-0\nonumber\\
    =&\bra{0}(\hat{p}e^{-r})^2\ket{0}\nonumber\\
    =&\frac{e^{-2r}}{2}.
\end{align}

Finally, we calculate the variance of the nullifer $\expval{[\Delta(\hat{p}_1-\hat{x}_2)]^2}+\expval{[\Delta(\hat{p}_2-\hat{x}_1)]^2}$ for the cluster state $\hat{C}_{\mathrm{z}}\ket{p=0}^{\otimes2}$. First, the term $\expval{[\Delta(\hat{p}_1-\hat{x}_2)]^2}$ is given by
\begin{equation}
\begin{aligned}
    &\expval{[\Delta(\hat{p}_1-\hat{x}_2)]^2}\nonumber\\
    =&\bra{\mathrm{sq}_{-r}}^{\otimes2}\hat{C}_{\mathrm{z}}^\dagger(\hat{p}_1-\hat{x}_2)^2\hat{C}_{\mathrm{z}}\ket{\mathrm{sq}_{-r}}^{\otimes2}\nonumber\\
    &-\left[\bra{\mathrm{sq}_{-r}}^{\otimes2}\hat{C}_{\mathrm{z}}^\dagger(\hat{p}_1-\hat{x}_2)\hat{C}_{\mathrm{z}}\ket{\mathrm{sq}_{-r}}^{\otimes2}\right]^2\nonumber\\
    =&\bra{\mathrm{sq}_{-r}}^{\otimes2}\hat{C}_{\mathrm{z}}^\dagger(\hat{C}_{\mathrm{z}}\hat{p}_1\hat{C}_{\mathrm{z}}^\dagger)^2\hat{C}_{\mathrm{z}}\ket{\mathrm{sq}_{-r}}^{\otimes2}\nonumber\\
    &-\left[\bra{\mathrm{sq}_{-r}}^{\otimes2}\hat{C}_{\mathrm{z}}^\dagger(\hat{C}_{\mathrm{z}}\hat{p}_1\hat{C}_{\mathrm{z}}^\dagger)\hat{C}_{\mathrm{z}}\ket{\mathrm{sq}_{-r}}^{\otimes2}\right]^2\nonumber\\
    =&\bra{\mathrm{sq}_{-r}}\hat{p}^2\ket{\mathrm{sq}_{-r}}\bra{\mathrm{sq}_{-r}}\ket{\mathrm{sq}_{-r}}\nonumber\\
    &-\bra{\mathrm{sq}_{-r}}\hat{p}\ket{\mathrm{sq}_{-r}}^2\bra{\mathrm{sq}_{-r}}
    \ket{\mathrm{sq}_{-r}}\nonumber\\
    =&\frac{e^{-2r}}{2}.
\end{aligned}
\end{equation}
In the same way, the term $\expval{[\Delta(\hat{p}_2-\hat{x}_1)]^2}$ turns into $e^{-2r}/2$. Hence, the variance of the nullifier for the cluster state can be calculated as 
\begin{equation}
\begin{aligned}
    &\expval{[\Delta(\hat{p}_1-\hat{x}_2)]^2}+\expval{[\Delta(\hat{p}_2-\hat{x}_1)]^2}\nonumber\\
    =&\frac{e^{-2r}}{2}+\frac{e^{-2r}}{2}\nonumber\\
    =&e^{-2r}.
\end{aligned}
\end{equation}

\section{Controlled squeezing and displacement using controlled phase shift}\label{sec:ControlledOperation}

\begin{figure}[t]
\centering
\includegraphics[width=1.\linewidth]{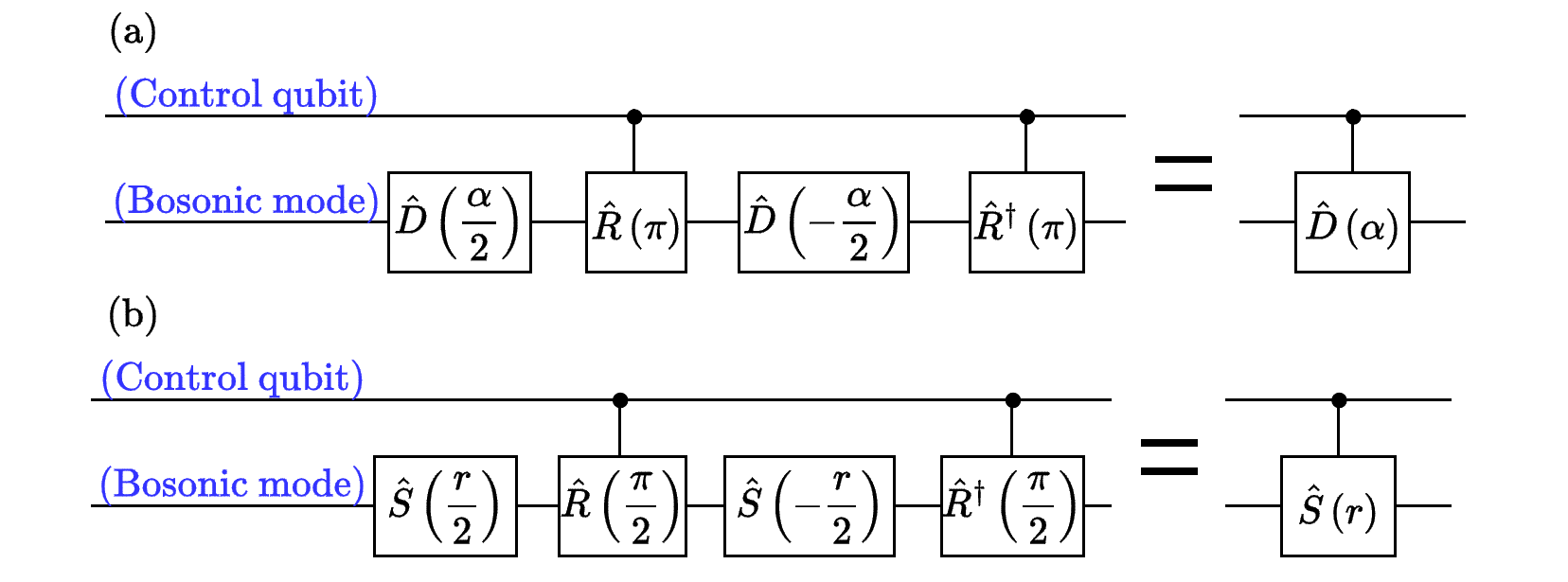}
\caption{Circuits to implement controlled operations with only controlled phase shift and single-mode gates. (a) Circuits to implement controlled displacement, proposed in Ref.~\cite{eickbusch2022fast}. (b) Circuits to implement controlled squeezing.}\label{fig:ControlledOperation}
\end{figure}
 While controlled phase shift and displacement operations are a standard technique in superconducting quantum circuits, controlled squeezing has not yet been experimentally realized and is expected to be more susceptible to noise. To address this, we propose circuits that implement controlled squeezing using only controlled phase shifts and single-mode squeezing gates~\cite{sivak2019kerr}. This proposal is inspired by the controlled displacement gate using the controlled phase shift gates and single-mode displacement gates, proposed in Ref.~\cite{eickbusch2022fast}.

First, we describe the realization of this controlled displacement operation~\cite{eickbusch2022fast} in Fig.~\ref{fig:ControlledOperation}(a). The left part of the circuit acts on the joint state of a control qubit and a bosonic mode as follows:
\begin{align}\label{eq:C-disp}
    &\ket{0}\bra{0}\otimes\left[\hat{I}\cdot\hat{D}\left(-\frac{\alpha}{2}\right)\cdot\hat{I}\cdot\hat{D}\left(\frac{\alpha}{2}\right)\right]\nonumber\\
    &+\ket{1}\bra{1}\otimes\left[\hat{R}^\dagger\left(\pi\right)\cdot\hat{D}\left(-\frac{\alpha}{2}\right)\cdot\hat{R}\left(\pi\right)\cdot\hat{D}\left(\frac{\alpha}{2}\right)\right]\nonumber\\
    &=\ket{0}\bra{0}\otimes\left[\hat{D}\left(-\frac{\alpha}{2}\right)\hat{D}\left(\frac{\alpha}{2}\right)\right]+\ket{1}\bra{1}\otimes\left[\hat{D}\left(\frac{\alpha}{2}\right)\hat{D}\left(\frac{\alpha}{2}\right)\right]\nonumber\\
    &=\ket{0}\bra{0}\otimes\hat{I}+\ket{1}\bra{1}\otimes\hat{D}(\alpha).
\end{align}
Hence, we show from Eq.~\eqref{eq:C-disp} that Fig.~\ref{fig:ControlledOperation}(a) acts as the controlled displacement.

Analogously, Fig.~\ref{fig:ControlledOperation}(b) implements a controlled squeezing operation. The relevant transformation is given by:
\begin{align}\label{eq:C-sq}
    &\ket{0}\bra{0}\otimes\left[\hat{I}\cdot\hat{S}\left(-\frac{r}{2}\right)\cdot\hat{I}\cdot\hat{S}\left(\frac{r}{2}\right)\right]\nonumber\\
    &+\ket{1}\bra{1}\otimes\left[\hat{R}^\dagger\left(\frac{\pi}{2}\right)\cdot\hat{S}\left(-\frac{r}{2}\right)\cdot\hat{R}\left(\frac{\pi}{2}\right)\cdot\hat{S}\left(\frac{r}{2}\right)\right]\nonumber\\
    &=\ket{0}\bra{0}\otimes\left[\hat{S}\left(-\frac{r}{2}\right)\hat{S}\left(\frac{r}{2}\right)\right]+\ket{1}\bra{1}\otimes\left[\hat{S}\left(\frac{r}{2}\right)\hat{S}\left(\frac{r}{2}\right)\right]\nonumber\\
    &=\ket{0}\bra{0}\otimes\hat{I}+\ket{1}\bra{1}\otimes\hat{S}(r).
\end{align}
Hence, we show from Eq.~\eqref{eq:C-sq} that Fig.~\ref{fig:ControlledOperation}(b) acts as the controlled squeezing.

This construction suggests that controlled squeezing can be achieved using only experimentally established components, which could potentially enable more robust implementations in current experimental technologies.

\section{Advantage of our projection onto cubic phase state}\label{sec:NoiseReduction_Projection}
\begin{figure}
    \centering
    \includegraphics[width=1.\linewidth]{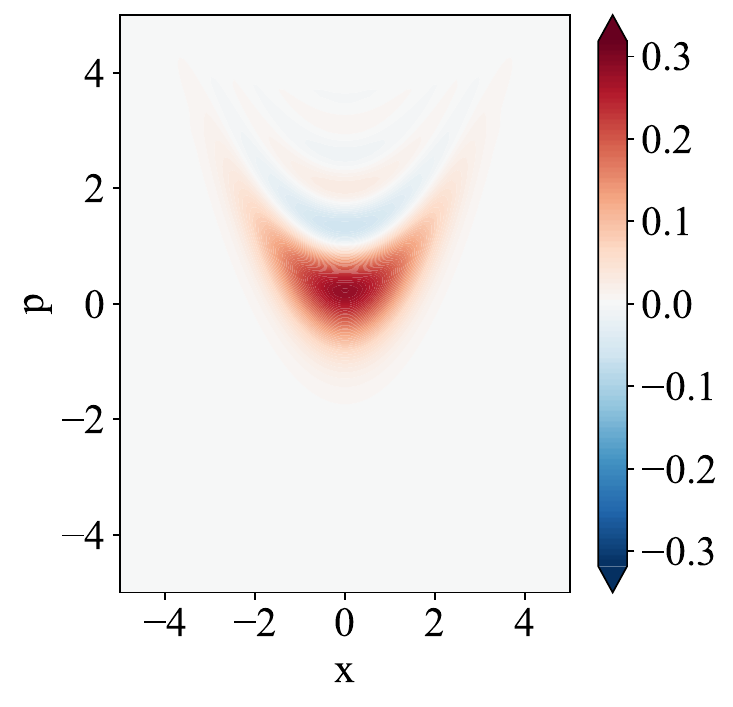}
    \caption{Wigner function of cubic phase state which is created from the vacuum state by utilizing only our projector.}
    \label{fig:comparison_for_ref2_a}
\end{figure}

\begin{figure*}
    \centering
    \includegraphics[width=0.95\linewidth]{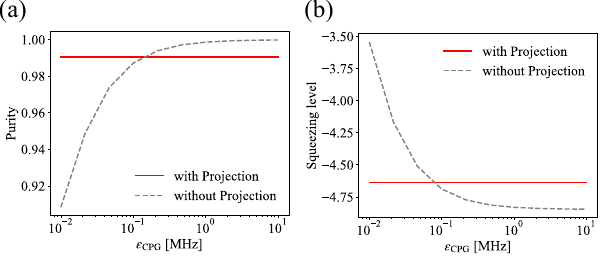}
    \caption{Comparison of the conventional method of cubic phase gate and our projective scheme. \red{(a) Evaluation with purity. (b) Evaluation with the squeezing level.}}
    \label{fig:comparison_for_ref2_b}
\end{figure*}

In this section, we numerically simulate the projection of the vacuum state onto a cubic phase state under photon loss noise when using the VQED method. \red{If one implements the cubic phase gate to generate a cubic phase state, third-order nonlinearities are required. In contrast, our projective scheme can be realized using only controlled phase-shift gates and other Gaussian operations, which rely at most on second-order nonlinearities. Consequently, our scheme is expected to be less demanding to implement than a direct cubic-gate realization. To examine this expectation,} we compare its performance with that of the conventional method, which directly implements the cubic phase gate to create the cubic phase state~\cite{hillmann2020universal}. This comparison reveals that our strategy is advantageous in a specific parameter regime. In numerical simulations, we use the vacuum state ($\eta = r = 0$) as the initial state for both the conventional method and our projective squeezing method, and aim to prepare a cubic phase state with $\eta = 0.3$ and $r = 5$ dB.

In the conventional method, we apply a squeezing operation described by the Hamiltonian
\begin{equation}
\hat{H}_{\mathrm{sq}} = \frac{\epsilon_{\mathrm{sq}}}{2}(\hat{x}\hat{p} + \hat{p}\hat{x}),
\label{eq:HamiltonianSq}
\end{equation}
followed by the application of the cubic phase gate via the Hamiltonian
\begin{equation}
\hat{H}_{\mathrm{CPG}} = -\frac{\epsilon_{\mathrm{CPG}}}{3}\hat{x}^3
\label{eq:HamiltonianCPG}
\end{equation}
to the resulting squeezed vacuum state. In the numerical simulation, the driving constant of the squeezing operation, $\epsilon_{\mathrm{sq}}$, is fixed at 10 MHz, which is of the same order as in Ref.~\cite{eriksson2024universal}. We vary the driving constant of the cubic phase gate, $\epsilon_{\mathrm{CPG}}$, from 10 kHz to 10 MHz in order to identify the parameter regimes where the cubic phase gate becomes experimentally challenging to implement, and where our protocol offers an advantage. The gate time for the squeezing operation is given by $t_{\mathrm{sq}} = r / \epsilon_{\mathrm{sq}} = 5~\mathrm{dB} / 10~\mathrm{MHz}$, while the gate time for the cubic phase gate is $t_{\mathrm{CPG}} = \eta / \epsilon_{\mathrm{CPG}} = 0.3 / \epsilon_{\mathrm{CPG}}$.

Now, we illustrate numerical simulations for cubic phase state generation using our projector when using the VQED method. We apply the projective squeezing projector defined in Eq.~\eqref{Eq: CPSpro} to the vacuum state to generate the cubic phase state. To implement this projector, we need to realize three types of operations: the controlled-displacement operation $\hat{D}(\alpha_\mathrm{CPG}(x_0))$, the controlled-phase shift operation $\hat{R}(\phi_\mathrm{CPG}(x_0))$, and the controlled-squeezing operation $\hat{S}(r_\mathrm{CPG}(x_0))$. Although the controlled-displacement and controlled-squeezing operations can be implemented using controlled-phase shifts and single-mode Gaussian operations,  we assume that each of them is implemented directly.

We model each operation using the following Hamiltonians:
\begin{align}
    \hat{H}_{\mathrm{c-disp}} &= \sqrt{2}\epsilon_{\mathrm{c-disp}}\ket{1}\bra{1}\otimes(\hat{x}\sin\theta_{\mathrm{c-disp}} - \hat{p}\cos\theta_{\mathrm{c-disp}})\nonumber\\
    &+\ket{0}\bra{0}\otimes\hat{I}, \nonumber\\
    \hat{H}_{\mathrm{c-phase}} &= \epsilon_{\mathrm{c-phase}} \ket{1}\bra{1}\otimes\hat{a}^\dagger \hat{a}+\ket{0}\bra{0}\otimes\hat{I}, \nonumber\\
    \hat{H}_{\mathrm{c-sq}} &= \frac{\epsilon_{\mathrm{c-sq}}}{2}\ket{1}\bra{1}\otimes(\hat{x} \hat{p} + \hat{p} \hat{x})+\ket{0}\bra{0}\otimes\hat{I},
    \label{eq:ControlledHamiltonians}
\end{align}
where the displacement angle $\theta_{\mathrm{c-disp}}$ is defined by
\[
\sin\theta_{\mathrm{c-disp}} = \frac{\Im(\alpha_\mathrm{CPG}(x_0))}{|\alpha_\mathrm{CPG}(x_0)|}, \quad 
\cos\theta_{\mathrm{c-disp}} = \frac{\Re(\alpha_\mathrm{CPG}(x_0))}{|\alpha_\mathrm{CPG}(x_0)|}.
\]

The driving constant for each operation is set to $\epsilon_{\mathrm{c-disp}} = \epsilon_{\mathrm{c-phase}} = \epsilon_{\mathrm{c-sq}} = 10\ \mathrm{MHz}$, which is of the same order as in Ref.~\cite{ths:ControlSq_1}. We estimate the typical magnitude of the gate parameter $x_0$ based on the standard deviation of the Gaussian distribution, using $x_0 \simeq 1/\sqrt{2\gamma}$. The parameter $\gamma$ is determined from Eq.~\eqref{eq:VirtualSq_2} and in our simulation, we set $r = 0\ \mathrm{dB}$ and $\Delta r = 5\ \mathrm{dB}$.

The gate parameters $\phi_{\mathrm{CPG},1}(x_0)$, $r_\mathrm{CPG}(x_0)$, $\phi_{\mathrm{CPG},2}(x_0)$, and $\alpha_\mathrm{CPG}(x_0)$ are given by Eq.~\eqref{eq:EvolutionTime}, where we set $\eta = 0.3$ in our simulation.

Using these expressions, the gate times for the first controlled-phase shift, controlled-squeezing, second controlled-phase shift, and controlled-displacement operations are estimated as follows:
\begin{align}
    t_{\mathrm{c-phase,1}} &= \frac{\phi_{\mathrm{CPG},1}(x_0)}{\epsilon_{\mathrm{c-phase}}}, \nonumber\\
    t_{\mathrm{c-sq}} &= \frac{r_\mathrm{CPG}(x_0)}{\epsilon_{\mathrm{c-sq}}}, \nonumber\\
    t_{\mathrm{c-phase,2}} &= \frac{\phi_{\mathrm{CPG},2}(x_0)}{\epsilon_{\mathrm{c-phase}}}, \nonumber\\
    t_{\mathrm{c-disp}} &= \frac{|\alpha_\mathrm{CPG}(x_0)|}{\epsilon_{\mathrm{c-disp}}}.
\end{align}

During the time evolution of both the conventional and our projective scheme, we account for realistic experimental noise by incorporating photon loss using the Lindblad master equation:
\begin{equation}
    \frac{\mathrm{d}\hat{\rho}}{\mathrm{d}t}=-i[\hat{H},\hat{\rho}]+\frac{\kappa}{2}(2\hat{a}\hat{\rho}\hat{a}^\dagger-\hat{a}^\dagger\hat{a}\hat{\rho}-\hat{\rho}\hat{a}^\dagger\hat{a}).\label{eq:lindblad}
\end{equation}
Here, we assume a damping rate of $\kappa = 5$ kHz during the whole process in our simulation, which is of the same order as in Ref.~\cite{ths:ControlSq_1}. Note that we do not account for the error in the ancillary qubit because this sort of error can be canceled for the VQED method~\cite{endo2024projective}. That is, we do not apply Hamiltonians in Eq.~\eqref{eq:ControlledHamiltonians} and instead, use it to estimate the gate time. We simulate our projective scheme by first applying our noiseless projector, and then, we consider the realistic noise in Eq.~\eqref{eq:lindblad} during the estimated gate time. In contrast, we directly apply Hamiltonians in Eqs.~\eqref{eq:HamiltonianSq},~\eqref{eq:HamiltonianCPG} for the conventional method.

Figure~\ref{fig:comparison_for_ref2_a} shows the Wigner function of the cubic phase state generated from the vacuum state using only our projector, without employing the cubic phase gate. This result numerically shows that our scheme can successfully prepare the cubic phase state without the need for a cubic phase gate, thereby ease the experimental challenges associated with implementing cubic or higher-order Hamiltonians.

Figure~\ref{fig:comparison_for_ref2_b} presents a comparison between the conventional method using direct implementation of the cubic phase gate and our projective scheme. The horizontal axis indicates the driving constant of the cubic phase gate $\epsilon_{\mathrm{CPG}}$, while the vertical axis in Fig.~\ref{fig:comparison_for_ref2_b}(a) \red{(Fig.~\ref{fig:comparison_for_ref2_b}(b))} represents the purity of the output states \red{(squeezing level $10\log_{10} \!\left\{2\,\expval{[\Delta (\hat{p}-\eta\hat{x}^2)]^2}\right\}$)}. The label ``with Projection'' refers to the results obtained using our projective method, whereas ``without Projection'' corresponds to those from the conventional method. From this comparison, we observe that our projective scheme outperforms the conventional method when the driving constant $\epsilon_{\mathrm{CPG}}$ is sufficiently low relative to the speed of other operations such as the controlled-phase shift.

Based on these numerical simulations, we draw two main conclusions in this section. First, our protocol is beneficial in situations where cubic or higher-order Hamiltonians are not available. Second, even when such Hamiltonians can be implemented, our scheme remains advantageous if their driving constants are significantly lower than those of more accessible operations, such as controlled-phase shift gates.

\end{document}